\documentclass{article}

\usepackage{iclr2027/iclr2027_conference,times}
\usepackage{amsmath,amssymb,amsfonts,bm}
\usepackage{booktabs}
\usepackage{multirow}
\usepackage[table]{xcolor}
\usepackage{graphicx}
\usepackage{placeins}
\usepackage{float}
\usepackage{microtype}
\usepackage{xspace}
\usepackage{hyperref}
\hypersetup{hidelinks}
\usepackage{url}

\definecolor{multibitrow}{RGB}{225,240,250}

\newcommand{\method}{\textsc{MultiBit}\xspace}

\title{Relevance-Resolution Transfer via Scale-Decomposable Fractional Diffusion for Multi-Length Cross-Modal Hash Retrieval}

\author{
\textbf{Xi Chen \quad Xu Chen \quad Xiangyang Jia \quad Ting Gan}\\
\textbf{Xu Zhang \quad Shuquan Wei \quad Sitong Fan}\\
School of Computer Science, Wuhan University\\
Wuhan, Hubei, China\\
\texttt{\{chenxi00233,xuchen,jxy,ganting\}@whu.edu.cn}\\
\texttt{\{zhangx0802,2021302111045,2026182110081\}@whu.edu.cn}
}

\iclrfinalcopy

\providecommand{\MBLang}[2]{#1}

\def\appendixNUSWIDETitle{Comparison on NUS-WIDE}
\def\appendixNUSWIDECaption{I2T and T2I mAP@all results (\%) of MultiBit and existing cross-modal hashing methods on NUS-WIDE with four hash-code lengths. MultiBit is shown in bold; the best fixed-length and the best multi-length baseline are underlined.}
\def\appendixTaskHeader{Task}
\def\appendixMethodHeader{Method}
\def\appendixReferenceHeader{Reference}
\def\appendixNUSWIDENote{The MultiBit row is shown in bold; within each column the best fixed-length method and the best multi-length method are underlined. I2T denotes image-to-text retrieval, whereas T2I denotes text-to-image retrieval.}
\def\appendixEfficiencyTitle{Performance--Efficiency Measurements}
\def\appendixEfficiencyCaption{Raw performance--efficiency measurements of representative methods on MIRFlickr-25K under the 128-bit setting.}
\def\appendixTrainableParamsHeader{Trainable Params (M) $\downarrow$}
\def\appendixFLOPsHeader{FLOPs (G) $\downarrow$}
\def\appendixTrainingTimeHeader{Training Time (h) $\downarrow$}
\def\appendixBestEpochHeader{Best Epoch $\downarrow$}
\def\appendixImageEncodingHeader{Image Encoding (ms) $\downarrow$}
\def\appendixTextEncodingHeader{Text Encoding (ms) $\downarrow$}
\def\appendixEfficiencyNote{$\downarrow$ indicates that lower is better; the best result in each column is shown in bold. All measurements follow the common CLIP ViT-B/32 backbone and 100-epoch training protocol described in Section~\ref{sec:experimental-setup}.}
\def\appendixDatasetHeader{Dataset}
\def\appendixResolutionDetailsTitle{Relevance Resolution Evaluation}
\def\appendixGradedRelevanceTitle{IDF-Weighted Shared-Label Evaluator.}
\def\appendixGradedRelevanceText{To score every method with the same external reference rather than with its own objective, all methods are evaluated against one shared IDF-weighted shared-label evaluator. Because this evaluator rewards rare labels through the same frequency prior that the relation teacher uses, the comparison is repeated against a second, frequency-free evaluator in Section~\ref{sec:appendix-decoupled-evaluator}. Here $\mathcal R$ is the retrieval candidate database, $R=|\mathcal R|$, and $n_c$ counts its items labeled with category $c$.}
\def\appendixTeacherQualityCaption{NDCG@100 of different relation teachers under the IDF-weighted shared-label evaluator and the label-name semantic evaluator.}
\def\appendixRelationTeacherHeader{Relation Teacher}
\def\appendixMetricDefinitionTitle{Metrics.}
\def\appendixMetricDefinitionText{For query $q$, $\pi_q$ is the Hamming ranking of the retrieval database $\mathcal R$, $\pi_q^*$ is the ideal ranking by descending $r_{qi}$, and $\mathcal P_q\subseteq\mathcal R$ is the common candidate pool for pairwise comparison. NDCG@100 assesses top-ranked utility; Kendall's $\tau_b$, Collision, and Inversion assess agreement, ties, and reversed orders among candidates with distinct graded relevance.}
\def\appendixResolutionProtocolTitle{Evaluation Protocol.}
\def\appendixResolutionProtocolText{Candidate identities are held fixed within each method comparison. NDCG@100 evaluates the first 100 Hamming-ranked candidates from $\mathcal R$; the pairwise measures use a common candidate pool $\mathcal P_q\subseteq\mathcal R$ for each query. Kendall's $\tau_b$ corrects for Hamming ties. Query metrics are averaged separately for I2T and T2I, and their arithmetic mean is reported in the resolution tables, where N denotes NDCG@100, $\tau_b$ tie-corrected Kendall's $\tau_b$, C the collision rate and I the inversion rate. The common pool $\mathcal P_q$ is the union of the top-100 candidates of the method being reported and of the full model, so $|\mathcal P_q|\le200$; the full model is therefore the common anchor of every row of the resolution tables, and its own row uses its own top-100 candidates. A candidate is relevant for mAP when it shares at least one label with the query, and Hamming ties are ordered by database index (stable sort) for every method. Equal-relevance pairs are excluded from Collision and Inversion, and equal-distance pairs count as collisions rather than inversions. Since every query has a relevant counterpart in the retrieval database, no query is dropped: the ideal NDCG denominator is always positive, and $\tau_b$ is 0 only when its tie-corrected denominator vanishes.}
\def\appendixParameterSensitivityTitle{FRT Hyperparameter Sensitivity}
\def\appendixParameterSensitivityCaption{Validation Mean mAP (\%) obtained with different combinations of the FRT hyperparameters $\alpha$ and $\eta$ on all datasets under the 64-bit setting.}
\begin{document}
\raggedbottom

\maketitle
\fancyhead{}

\begin{abstract}
{\looseness=-1
Cross-modal hashing enables efficient retrieval by encoding heterogeneous data into compact binary codes. Recent methods exploit fine-grained relations encoded in multi-label training structure, yet none of them constrains how those relations survive as consistent candidate rankings in finite, multi-length Hamming spaces, which we term the relevance resolution bottleneck (RRB). To address the RRB, we propose MultiBit, which transfers relevance resolution from multi-label structure to multi-length Hamming spaces. MultiBit first constructs a scale-decomposable fractional relation teacher from dataset-level label co-occurrence and label specificity, and models dependencies from local to long-range over continuous diffusion scales. It then maps the discretized diffusion scales and their quadrature weights to scale-aware bit subblocks of the maximum-length code, organizes the target code lengths as nested prefixes, and aligns their Hamming candidate rankings with the teacher relations. Experiments on multiple benchmarks demonstrate improved retrieval accuracy. \emph{Code is available in the supplementary material.}\par}
\end{abstract}

\section{Introduction}
\label{sec:introduction}

Cross-modal hashing maps heterogeneous image and text data into a shared Hamming space, enabling efficient retrieval through compact storage and bitwise comparison~\citep{wang2024crossmodalreview}. At inference, a query is encoded from its observed modality and ranked against the database codes by Hamming distance. Retrieval quality therefore depends not only on separating relevant from irrelevant samples, but also on whether semantic differences among candidates are reflected in their relative distances.

Recent methods increasingly exploit fine-grained relations encoded in multi-label training structure, from hard-negative objectives that separate relevant from irrelevant samples~\citep{zhang2026hardnegatives,su2025neighboraware,peng2026semantic}, through proxy- and overlap-based objectives that distinguish degrees of observed agreement~\citep{huo2024hierarchy,yao2026compositeproxy,su2026ambiguity}, to structured-label approaches that capture dataset-level dependencies and already produce continuous relation scores~\citep{cao2025deepgraph,zhang2026adaptive,meng2026scenegraph}. These developments progressively improve label-side resolution, yet a fine-grained relation in label space does not by itself determine how propagation ranges should be encoded in finite bits, nor does it guarantee that the induced order survives in the final Hamming list or across code lengths. The remaining limitation lies less in relation construction than in transferring these relations into finite, multi-length codes.

This transfer gap exposes a relevance resolution bottleneck (RRB): relevance resolution denotes the degree to which semantically grounded differences between a query and its candidates can be distinguished and preserved, rather than the number of bits itself. When label-side relations enter a finite Hamming space without explicit organization and ranking constraints, candidates with different relevance can collide at the same distance, appear in inverted order, or exchange positions across code lengths. The resulting question is how to organize such relations as transferable supervision that is preserved consistently in nested, multi-length Hamming rankings.

\begin{figure*}[t]
	\centering
	\includegraphics[width=\textwidth]{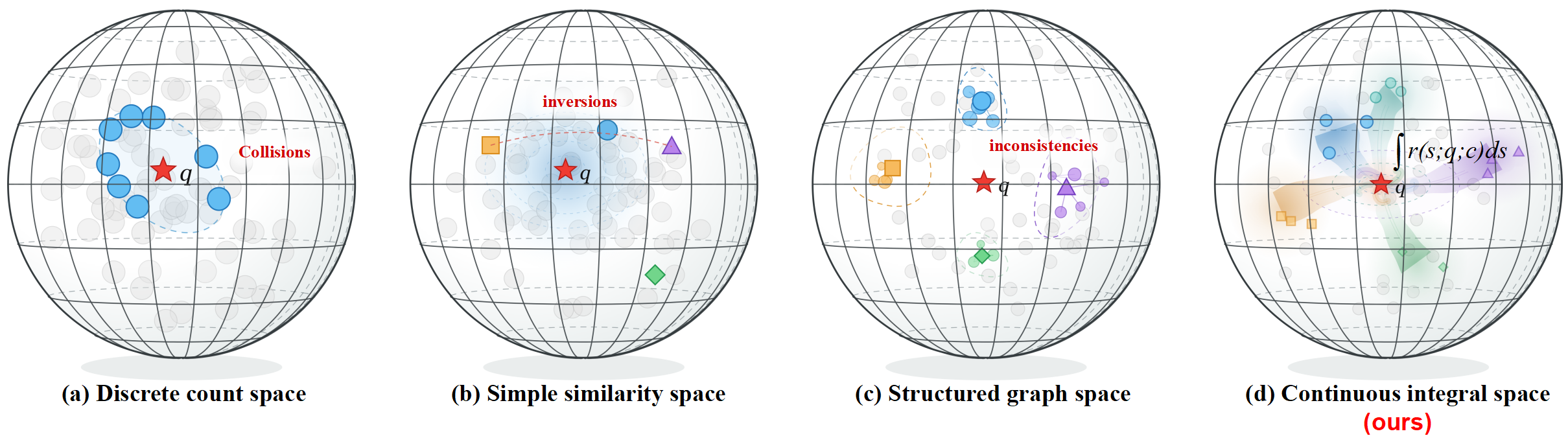}
	\caption{Motivation for multi-scale relation modeling. (a)--(c) illustrate relation spaces built from discrete labels, simple similarity, and graph structure, which capture progressively richer relations but still describe them at a single scale; (d) MultiBit integrates label co-occurrence and label specificity through fractional diffusion, covering dependencies from local neighborhoods to long-range graph paths and exposing the resulting scale structure to finite binary codes.}
	\label{fig:motivation}
\end{figure*}

As illustrated in Figure~\ref{fig:motivation}, RRB manifests as distance collisions and ranking inversions among semantically different candidates; its cross-length form is quantified in Section~\ref{sec:multi-length-analysis}. To address it, we propose \method, a unified framework that transfers relevance resolution from dataset-level multi-label structure to multi-length Hamming spaces. The framework starts from a scale-decomposable Fractional Relation Teacher (FRT), which integrates label co-occurrence and label specificity through fractional diffusion and decomposes the result into query-conditioned relevance targets together with the quadrature weight of each diffusion scale. These weights then guide Scale-aware Bit Organization (SBO), which assigns relation scales to dedicated bit subblocks of the maximum-length code, and Multi-length Prefix Alignment (MPA), which organizes the target code lengths as nested prefixes and constrains their Hamming candidate distributions to follow the same teacher-induced ordering.
Our contributions are threefold:
\begin{itemize}
  \item We identify and formalize the relevance resolution bottleneck (RRB), in which label-side relations collide, invert, or exchange ranks across code lengths.
  \item We propose \method, which constructs relevance resolution with FRT and preserves it in nested multi-length Hamming spaces through SBO and MPA.
  \item We characterize a conditional order-preservation property of the listwise objective in the continuous Hamming surrogate, and evaluate how far that ordering survives sign quantization at all four code lengths.
\end{itemize}

\section{Related Work}
\label{sec:related-work}

\noindent\textbf{Cross-Modal Hashing.} Existing cross-modal hashing methods can be grouped into three categories by how they construct semantic supervision. The first employs pairwise, contrastive, triplet, or hard-negative objectives on binary relevance, establishing a basic discriminative boundary~\citep{zhang2026hardnegatives,su2025neighboraware,peng2026semantic}. The second introduces multi-label classification, semantic proxies, shared-label counts, or Jaccard similarity to retain label identities and direct overlap~\citep{huo2024hierarchy,yao2026compositeproxy,cheng2026distanceweighted}. The third incorporates structured label relations through label graphs, propagation, embeddings, or semantic hierarchies, extending isolated indicators with dataset-level co-occurrence and higher-order dependencies~\citep{zhang2026dualgraph,cao2025deepgraph,zhang2026adaptive,meng2026scenegraph}.

\noindent\textbf{Fine-Grained Relevance Modeling.} Fine-grained relevance modeling describes graded relations that a single relevant/irrelevant indicator cannot express. Direct set-based measures distinguish candidates by observed label overlap, while frequency-aware weighting reduces the dominance of common labels~\citep{yang2025similarity}; structured approaches further exploit label correlations, graph neighborhoods, and indirect semantic paths, and some produce continuous relation scores for pairwise supervision or representation regularization~\citep{su2026ambiguity,cheng2026distanceweighted}.

\noindent\textbf{Multi-Length and Bit-Scalable Hashing.} A separate line of work targets deploying a single model across several code lengths. Central-coding and multi-bit schemes share one representation across lengths~\citep{wu2024cmcl,zuo2026beyond}, and adaptive-bit methods allocate capacity across bits or lengths~\citep{wang2025adaptivebit,zou2025prompthash,jiang2026regenerated}. These approaches reduce deployment cost, but they do not constrain how the induced candidate ranking changes when a code is truncated.

Despite these advances, existing methods focus on constructing stronger semantic relations or improving feature discrimination; the resulting relations are not organized according to propagation scale and finite bit capacity, and are not constrained to survive as consistent candidate rankings across code lengths~\citep{zuo2026beyond,tan2026bitwise,jiang2026regenerated}. Label-side distinctions may therefore still collapse into distance ties, ranking inversions, or cross-length inconsistencies after binary encoding. To address this, \method\ couples FRT, SBO, and MPA to preserve fine-grained semantic relations as consistent candidate rankings in multi-length Hamming spaces.

\section{Method}
\label{sec:method}

\subsection{Problem Formulation: Relevance Resolution Bottleneck}
\label{sec:problem}

Let $\mathcal{D}=\{(x_i^I,x_i^T,y_i)\}_{i=1}^{N}$ denote a cross-modal training set, where $x_i^I$ and $x_i^T$ are the image and text views of the $i$-th instance, respectively, and $y_i\in\{0,1\}^{C}$ is their shared $C$-dimensional multi-hot label vector, with $N$ denoting the number of training instances. Given a set of target code lengths $\mathcal{B}=\{B_1,\ldots,B_K\}$ and its maximum length $B_{\max}=\max\mathcal{B}$, we learn modality-specific hash functions $F^I$ and $F^T$ that map heterogeneous inputs into a shared $B_{\max}$-dimensional Hamming space. Since the discrete sign function is non-differentiable, we follow the continuous-relaxation strategy used in deep hashing~\citep{zou2025prompthash,tan2026bitwise} and employ $\tanh$ as a continuous relaxation during training, so that the maximum-length representation of instance $i$ in modality $m$ is $\widetilde h_i^m=\tanh(F^m(x_i^m))\in(-1,1)^{B_{\max}}$. To support multiple code lengths with a single model, as in multi-bit and unequal-length hashing~\citep{wu2024cmcl,zuo2026beyond,zou2025prompthash,jiang2026regenerated,wang2025adaptivebit}, we define a prefix selection matrix $P_B=[I_B\ \mathbf{0}]\in\{0,1\}^{B\times B_{\max}}$ and obtain the binary code at any target length $B\in\mathcal{B}$ from the first $B$ dimensions of the maximum-length representation as $b_i^{m,(B)}=\operatorname{sign}(P_B\widetilde h_i^m)\in\{-1,+1\}^{B}$. Accordingly, the Hamming distance between an image query $x_i^I$ and a text candidate $x_j^T$ at code length $B$ is
\begin{equation}
d_H^{(B)}(i,j)
=
\frac{1}{2}
\left(
B-
{b_i^{I,(B)}}^{\mathsf T}b_j^{T,(B)}
\right),
\label{eq:hamming-distance}
\end{equation}
with a symmetric formulation for text-to-image retrieval. Labels are used only to construct training supervision; inference requires only encoding, prefix truncation, and Hamming ranking. Appendix Section~\ref{sec:appendix-rrb} states RRB formally, in its collision, inversion, and cross-length forms.

\subsection{Overview}
\label{sec:overview}

\begin{figure*}[t]
  \centering
  \includegraphics[width=\textwidth]{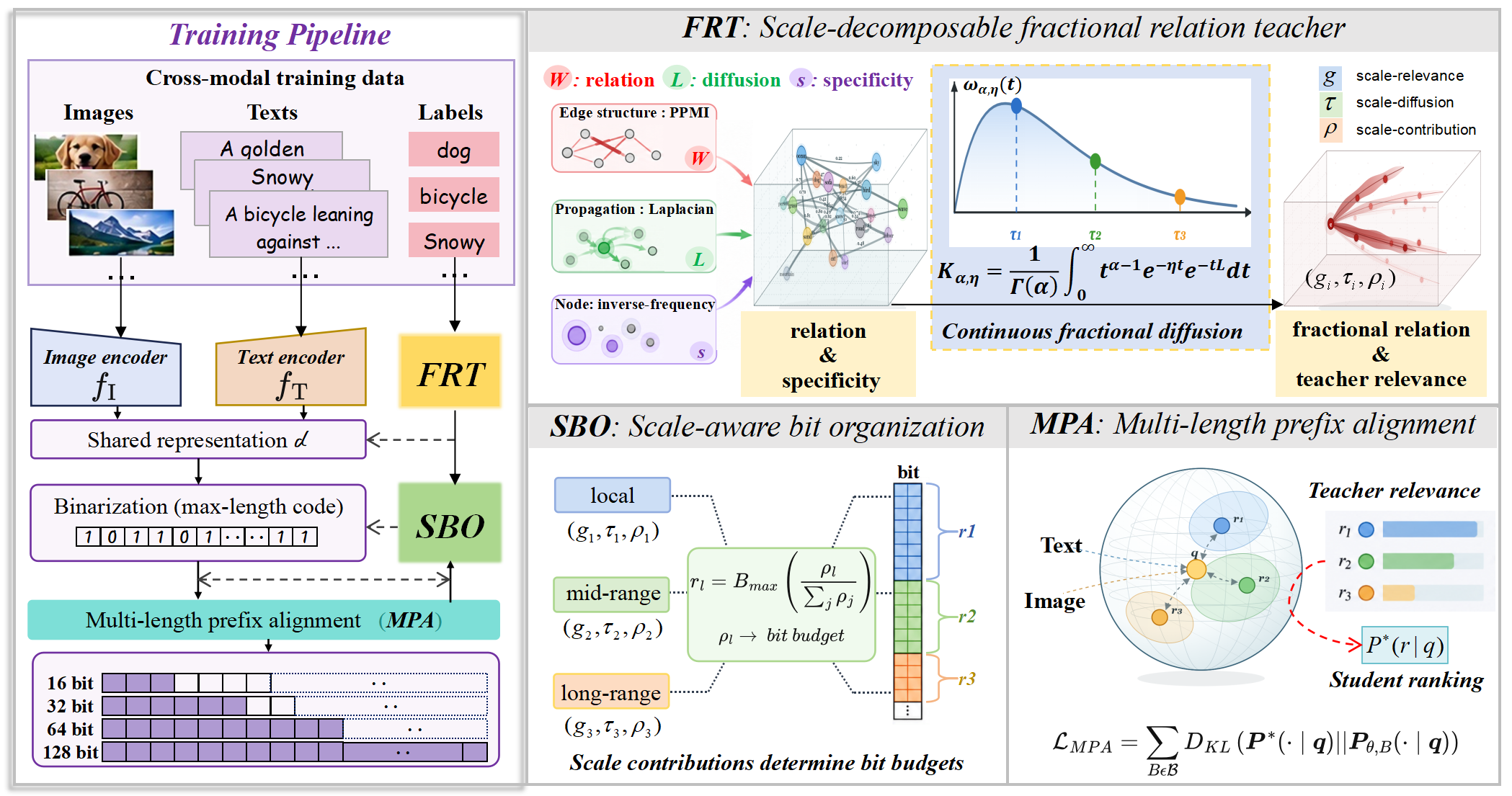}
\caption{
	Overview of \method. FRT derives an integrated teacher $g^*$ through fractional diffusion and decomposes it into $M$ scale components $\{(g_\ell,\tau_\ell,\rho_\ell)\}$. SBO converts $\rho_\ell$ into bit budgets $m_\ell$ within $B_{\max}$, and MPA aligns the Hamming distributions $P_{\theta,B}$ of nested prefixes with the teacher $P^*$ at every target length.
}
  \label{fig:framework}
\end{figure*}

As illustrated in Figure~\ref{fig:framework}, \method\ consists of three components; the notation and experimental settings are summarized in Appendix Table~\ref{tab:all-parameters}. FRT turns training-set label co-occurrence and specificity into continuous local-to-long-range dependencies and decomposes the resulting teacher into finite diffusion scales with their quadrature weights. SBO allocates finite representation capacity by mapping these scales and weights to scale-specific bit subblocks. MPA organizes the subblocks as nested prefixes and aligns their Hamming rankings with one shared teacher ordering.

\subsection{Scale-Decomposable Fractional Relation Teacher (FRT)}
\label{sec:teacher}

The first component constructs the relational supervision that the Hamming representations should preserve. Given the training labels defined in Section~\ref{sec:problem}, let $n_a=\sum_i y_{ia}$ and $n_{ab}=\sum_i y_{ia}y_{ib}$ denote the frequency of label $a$ and the co-occurrence frequency of labels $a$ and $b$, respectively. We build a positive pointwise mutual information (PPMI) label graph $W\in\mathbb{R}^{C\times C}$~\citep{levy2014neural,hou2025noisy} with a smoothing constant $\varepsilon>0$, whose entry $W_{ab}$ is the positive part of the mutual information between labels $a$ and $b$ (Appendix Section~\ref{sec:appendix-derivations}, Eq.~(\ref{eq:ppmi-graph})). With $D_{aa}=\sum_bW_{ab}$, the normalized graph Laplacian is $L=I-D^{-1/2}WD^{-1/2}$. We further assign each label a specificity weight $s_a=\log((N+1)/(n_a+1))$ and obtain the weighted label vector $\widetilde y_i=y_i\odot s$; the resulting graph captures dataset-level label dependencies while $s$ modulates the contribution of each label by its frequency.

A single propagation depth describes only one semantic neighborhood. To integrate dependencies from direct co-occurrence to long-range graph paths, we build on heat-kernel diffusion over graphs~\citep{chamberlain2021grand} and define the continuous-scale fractional diffusion kernel
\begin{equation}
K_{\alpha,\eta}
=(L+\eta I)^{-\alpha}
=
\frac{1}{\Gamma(\alpha)}
\int_{0}^{\infty}
t^{\alpha-1}e^{-\eta t}e^{-tL}\,\mathrm{d}t,
\qquad \alpha>0,\ \eta>0.
\label{eq:fractional-kernel}
\end{equation}
Here $t$ is the diffusion scale, and $\alpha$ and $\eta$ control the spectral decay and the diffusion range, respectively~\citep{maskey2023fractional}; their interpretation is detailed in Appendix Section~\ref{sec:appendix-derivations}. For a query $q$ and a candidate $r$, the integrated teacher relevance is defined by the kernel-induced cosine
\begin{equation}
g^*(q,r)
=
\frac{\widetilde y_q^{\mathsf T}K_{\alpha,\eta}\widetilde y_r}
{\sqrt{
(\widetilde y_q^{\mathsf T}K_{\alpha,\eta}\widetilde y_q)
(\widetilde y_r^{\mathsf T}K_{\alpha,\eta}\widetilde y_r)}}.
\label{eq:integrated-teacher}
\end{equation}

The continuous kernel must expose a finite scale structure before it can guide finite bits. After the change of variable $x=\eta t$, we apply an $M$-point generalized Gauss--Laguerre rule~\citep{golub1969calculation} whose nodes $x_\ell$ are the roots of $L_M^{(\alpha-1)}(x)$ and whose weights are $\omega_\ell$. This yields
\begin{equation}
K_{\alpha,\eta}
\approx
K_{\alpha,\eta}^{(M)}
=
\sum_{\ell=1}^{M}\rho_\ell K_\ell,
\qquad
K_\ell=e^{-\tau_\ell L},\quad
\tau_\ell=\frac{x_\ell}{\eta},\quad
\rho_\ell=\frac{\eta^{-\alpha}\omega_\ell}{\Gamma(\alpha)}.
\label{eq:quadrature-decomposition}
\end{equation}
The rule supplies only the scale coordinates $\{(\tau_\ell,\rho_\ell)\}$; $g^*$ and every $g_\ell$ below are computed from the exact matrix functions $(L+\eta I)^{-\alpha}$ and $e^{-\tau_\ell L}$ (Appendix Table~\ref{tab:quadrature-fidelity}).
Each component also induces a scale-specific teacher
\begin{equation}
g_\ell(q,r)
=
\frac{\widetilde y_q^{\mathsf T}K_\ell\widetilde y_r}
{\sqrt{
(\widetilde y_q^{\mathsf T}K_\ell\widetilde y_q)
(\widetilde y_r^{\mathsf T}K_\ell\widetilde y_r)}},
\qquad \ell=1,\ldots,M.
\label{eq:scale-teacher}
\end{equation}
FRT therefore outputs the integrated teacher ranking $g^*$ and the finite set $\{(g_\ell,\tau_\ell,\rho_\ell)\}_{\ell=1}^{M}$, which exposes the local-to-long-range scale structure to the subsequent bit organization.

\subsection{Scale-Aware Bit Organization (SBO)}
\label{sec:bit-organization}

The quadrature weights in Section~\ref{sec:teacher} describe how the fractional kernel distributes its mass over diffusion scales, and we use them as a prior for allocating finite representation capacity. For the maximum code length $B_{\max}$ defined in Section~\ref{sec:problem}, the ideal number of bits assigned to scale $\ell$ is
\begin{equation}
m_\ell^{\mathrm{ideal}}
=
B_{\max}
\frac{\rho_\ell}{\sum_{j=1}^{M}\rho_j}.
\label{eq:bit-budget}
\end{equation}

Since $\rho_\ell=\eta^{-\alpha}\omega_\ell/\Gamma(\alpha)$, the common factor cancels in the normalized ratio, so the allocation is governed by the Gauss--Laguerre weights $\{\omega_\ell\}$, hence by $\alpha$ rather than $\eta$. This ratio is a kernel-mass prior over scales rather than a measured per-scale contribution (Appendix Section~\ref{sec:appendix-derivations}).
Because bit dimensions must be integers, we convert $m_\ell^{\mathrm{ideal}}$ to $m_\ell$ using a largest-remainder allocation, reserve at least one bit for every retained scale, and adjust the remaining dimensions such that $\sum_{\ell=1}^{M}m_\ell=B_{\max}$. Thus, scales carrying more quadrature mass receive more capacity, while low-weight long-range components still retain a share of it; bit-scalable and unequal-length hashing allocate capacity in the same spirit~\citep{wang2025adaptivebit,zou2025prompthash}.

Let $E^m$ be the encoder and projection for modality $m\in\{I,T\}$, mapping an input to a shared $d$-dimensional feature $z_i^m=E^m(x_i^m)$. For each diffusion scale, a shared scale-specific hash head $H_\ell:\mathbb{R}^{d}\rightarrow\mathbb{R}^{m_\ell}$ produces a continuous subblock. We sort the scales from local to long range, denoted by $(\sigma_1,\ldots,\sigma_M)=\operatorname{argsort}(\tau_1,\ldots,\tau_M)$, and concatenate the resulting subblocks in that order to form the maximum-length representation (Appendix Section~\ref{sec:appendix-derivations}, Eq.~(\ref{eq:scale-subblocks})). This refines the hash function $F^m$ introduced in Section~\ref{sec:problem}: the maximum-length representation is no longer an unstructured vector, but a concatenation of subblocks with explicit diffusion-scale semantics. During training, each subblock is paired with its corresponding scale teacher $g_\ell$; the resulting scale-specific supervision is jointly defined with prefix alignment in the next subsection.

\subsection{Multi-Length Prefix Alignment (MPA)}
\label{sec:prefix-alignment}

Scale-aware subblocks specify the internal semantics of the maximum-length code, the only deployable length that is itself multi-scale (Table~\ref{tab:scale-coverage}). Consistent rankings across truncations are obtained by aligning the prefix rankings with the shared teacher ordering. All lengths share the same prefix operator, encoder, scale coordinates and bit ordering, so extending a code preserves every bit of its shorter prefix.

For a query $q$ and its cross-modal candidate set $\mathcal C_q$, the integrated teacher score from Eq.~(\ref{eq:integrated-teacher}) is converted into a query-conditioned teacher distribution $P^*(r\mid q)\propto\exp(g^*(q,r)/T_t)$ over $r\in\mathcal C_q$, where $T_t>0$ controls the concentration of the teacher ranking without changing its order~\citep{hwang2025tcil}. At length $B$, we compute a normalized differentiable Hamming surrogate and the corresponding student distribution for retrieval direction $m\rightarrow n$ as
\begin{equation}
\begin{aligned}
\overline d_{B}^{m\rightarrow n}(q,r)
&=\frac{1}{2B}
\left(
B-
(\widetilde h_q^{m,(B)})^{\mathsf T}
\widetilde h_r^{n,(B)}
\right),\\
P_{\theta,B}^{m\rightarrow n}(r\mid q)
&=\frac{\exp\!\left(-\overline d_B^{m\rightarrow n}(q,r)/T_s\right)}
{\sum_{u\in\mathcal C_q}
\exp\!\left(-\overline d_B^{m\rightarrow n}(q,u)/T_s\right)},
\end{aligned}
\label{eq:student-distribution}
\end{equation}
where $T_s>0$ is the student temperature and division by $B$ places different code lengths in the common range $[0,1]$. For a query set $\mathcal Q_m$, the directional listwise loss~\citep{yoon2025acurank} minimizes the Kullback--Leibler divergence~\citep{kullback1951information} between the two distributions; we denote the resulting loss at length $B$ by $\mathcal L_B^{m\rightarrow n}$ (Appendix Section~\ref{sec:appendix-derivations}, Eq.~(\ref{eq:directional-list})). The multi-length prefix objective aligns both retrieval directions at every target length:
\begin{equation}
\mathcal L_{\mathrm{prefix}}
=
\sum_{B\in\mathcal B}\beta_B
\left(
\mathcal L_B^{I\rightarrow T}
+\mathcal L_B^{T\rightarrow I}
\right),
\qquad \beta_B\ge0.
\label{eq:prefix-loss}
\end{equation}

We additionally apply the same bidirectional distribution matching to every complete scale subblock, replacing $g^*$ with $g_\ell$ and the length-$B$ prefix with $\widetilde h^{m,(\ell)}$. Let the resulting losses be $\mathcal L_\ell^{I\rightarrow T}$ and $\mathcal L_\ell^{T\rightarrow I}$. With normalized scale weights $\bar\rho_\ell=\rho_\ell/\sum_j\rho_j$, the scale objective and the complete training objective are
\begin{equation}
\begin{aligned}
\mathcal L_{\mathrm{scale}}
&=\sum_{\ell=1}^{M}\bar\rho_\ell
\left(
\mathcal L_\ell^{I\rightarrow T}
+\mathcal L_\ell^{T\rightarrow I}
\right),\\
\mathcal L
&=\mathcal L_{\mathrm{prefix}}
+\lambda_s\mathcal L_{\mathrm{scale}}.
\end{aligned}
\label{eq:overall-objective}
\end{equation}
Scale supervision is designed to make each subblock encode its designated propagation range, whereas prefix supervision aligns the continuous ranking at every deployable length with the integrated teacher order. In particular, if the listwise divergence at a given length vanishes, equality of the two distributions and the monotonicity of Softmax imply
\begin{equation}
g^*(q,r_a)>g^*(q,r_b)
\quad\Longrightarrow\quad
\overline d_B(q,r_a)<\overline d_B(q,r_b),
\label{eq:order-preservation}
\end{equation}
which links the teacher's relevance resolution to the student's soft Hamming ranking during training. Appendix Section~\ref{sec:appendix-quantization} and Section~\ref{sec:resolution-analysis} report how much of this order survives binarization and how well the discrete ranking preserves it. At inference the teacher branch and the listwise objectives are removed, so any requested length $B$ is served by prefix extraction and standard Hamming ranking.

\section{Experiments}
\label{sec:experiments}

\subsection{Datasets and Experimental Setup}
\label{sec:experimental-setup}

\noindent We conduct experiments on three public multi-label image--text datasets: MIRFlickr-25K~\citep{huiskes2008mirflickr}, MS-COCO~\citep{lin2014coco}, and IAPR TC-12~\citep{grubinger2006iapr}, and report additional results on NUS-WIDE~\citep{chua2009nuswide} in Appendix Table~\ref{tab:nuswide-comparison}. All methods use the same CLIP ViT-B/32 backbone~\citep{radford2021clip}. For the data split and the mAP@all metric we report image-to-text (I2T) and text-to-image (T2I) retrieval and follow the protocol of \citet{cheng2026dghdgh}; Appendix Section~\ref{sec:appendix-setup} gives dataset sizes and metrics.

\subsection{Comparison with State-of-the-Art Methods}
\label{sec:sota-comparison}

\noindent\textbf{Compared Methods.} We compare MultiBit with eight fixed-length cross-modal hashing methods (DSSH, DNPH, DECH, DIMCH, DPBE, DSTH, DPSIH, DGHDGH~\citep{qin2026dssh,huo2024dnph,li2025dech,tu2025dimch,cheng2025dpbe,qin2026dsth,han2026dpsih,cheng2026dghdgh}) and three multi-length methods serving several lengths from one model (CMCL~\citep{wu2024cmcl}, RCMH~\citep{jiang2026regenerated}, VH-ABL~\citep{zuo2026beyond}).

\begin{table}[htbp]\centering
\caption{I2T and T2I mAP@all (\%) on three benchmarks at four code lengths; best in bold, second best underlined.}
\label{tab:sota-comparison}
\scriptsize
\setlength{\tabcolsep}{2.4pt}
\renewcommand{\arraystretch}{0.90}
\resizebox{\textwidth}{!}{%
\begin{tabular}{cllcccccccccccc}
\toprule
\textbf{Task} & \textbf{Method} & \textbf{Reference}
& \multicolumn{4}{c}{\textbf{MIRFlickr-25K}}
& \multicolumn{4}{c}{\textbf{IAPR TC-12}}
& \multicolumn{4}{c}{\textbf{MS-COCO}} \\
\cmidrule(lr){4-7}\cmidrule(lr){8-11}\cmidrule(lr){12-15}
& & & \textbf{16} & \textbf{32} & \textbf{64} & \textbf{128}
& \textbf{16} & \textbf{32} & \textbf{64} & \textbf{128}
& \textbf{16} & \textbf{32} & \textbf{64} & \textbf{128} \\
\midrule
\multirow{8}{*}{\textbf{I2T}}
& DSSH   & TKDE'26  & 75.17 & 75.59 & 80.47 & 81.10 & 44.09 & 53.73 & 57.00 & 65.21 & 62.85 & 67.74 & 73.24 & \underline{76.38} \\
& DNPH   & TOMM'24  & 74.12 & 79.10 & 81.36 & 81.70 & 46.13 & 49.68 & 56.07 & 62.57 & 64.22 & 70.30 & 71.61 & 72.52 \\
& DECH   & AAAI'25  & 81.77 & 83.85 & 85.00 & \underline{85.44} & 58.54 & \underline{64.53} & \underline{67.65} & 69.47 & 64.20 & 68.70 & 72.39 & 74.06 \\
& DIMCH  & TIP'25   & 81.15 & 82.25 & 82.45 & 83.35 & \underline{58.73} & 61.93 & 64.23 & 65.48 & 65.18 & 67.65 & 69.80 & 69.33 \\
& DPBE   & MM'25    & 69.31 & 72.14 & 74.09 & 75.80 & 49.86 & 55.94 & 60.98 & 64.65 & 55.42 & 61.78 & 66.92 & 72.39 \\
& DSTH   & TMM'26   & 79.67 & 81.01 & 82.43 & 82.85 & 57.89 & 62.06 & 64.73 & 66.78 & 62.81 & 69.00 & 70.89 & 72.51 \\
& DPSIH  & AAAI'26  & 77.43 & 80.76 & 81.90 & 82.40 & 56.55 & 60.88 & 64.26 & 66.38 & 63.49 & 68.05 & 70.90 & 72.47 \\
& DGHDGH & ICLR'26  & 79.14 & 80.47 & 80.99 & 81.72 & 55.44 & 59.87 & 62.47 & 63.86 & 58.29 & 61.29 & 60.44 & 63.41 \\
\arrayrulecolor{gray!60}\cmidrule[0.4pt]{1-15}\arrayrulecolor{black}
\multirow{4}{*}{\shortstack{\textbf{I2T}\\\textbf{(Multi/length)}}}
& CMCL   & TKDE'24  & \textbf{83.13} & \underline{84.34} & \underline{85.22} & \underline{85.44} & 58.02 & 62.69 & 66.91 & \underline{69.59} & \textbf{70.23} & \textbf{72.95} & \underline{75.13} & 75.67 \\
& RCMH   & TPAMI'26 & 70.80 & 71.61 & 72.10 & 72.45 & 56.90 & 59.49 & 60.56 & 61.05 & 67.82 & 70.79 & 73.06 & 75.00 \\
& VH-ABL & TMM'26   & 80.28 & 81.73 & 82.51 & 82.74 & 54.71 & 60.31 & 63.60 & 65.37 & 63.32 & 70.35 & 73.61 & 75.38 \\
\rowcolor{multibitrow}
& \textbf{MultiBit} & \textbf{OURS} & \underline{82.86} & \textbf{84.37} & \textbf{85.49} & \textbf{85.92} & \textbf{60.95} & \textbf{65.27} & \textbf{68.25} & \textbf{70.07} & \underline{69.55} & \underline{72.53} & \textbf{75.68} & \textbf{76.89} \\
\midrule
\multirow{8}{*}{\textbf{T2I}}
& DSSH   & TKDE'26  & 74.29 & 73.63 & 78.65 & 79.38 & 43.17 & 51.10 & 55.91 & 68.36 & 63.26 & 68.48 & 74.66 & \underline{77.68} \\
& DNPH   & TOMM'24  & 71.96 & 76.43 & 78.41 & 79.18 & 44.28 & 48.13 & 56.04 & 63.62 & 64.35 & 70.05 & 71.05 & 72.61 \\
& DECH   & AAAI'25  & 78.11 & 79.86 & \underline{80.54} & \underline{81.09} & 57.41 & \underline{64.89} & 68.76 & 71.05 & 64.66 & 69.66 & 72.55 & 74.63 \\
& DIMCH  & TIP'25   & 76.89 & 77.60 & 79.01 & 78.76 & 58.28 & 61.96 & 64.49 & 65.87 & 64.77 & 67.51 & 69.31 & 68.54 \\
& DPBE   & MM'25    & 71.55 & 73.57 & 75.52 & 76.68 & 48.30 & 54.13 & 59.01 & 62.73 & 56.05 & 61.87 & 67.02 & 72.81 \\
& DSTH   & TMM'26   & 75.64 & 77.60 & 78.62 & 79.24 & 57.93 & 61.98 & 65.23 & 67.14 & 63.39 & 68.85 & 70.83 & 72.87 \\
& DPSIH  & AAAI'26  & 73.92 & 77.39 & 78.09 & 78.60 & 56.66 & 60.62 & 64.91 & 66.88 & 63.48 & 68.09 & 70.92 & 72.71 \\
& DGHDGH & ICLR'26  & 76.27 & 77.61 & 78.39 & 78.93 & 54.14 & 58.38 & 60.66 & 62.36 & 58.61 & 61.21 & 62.02 & 63.53 \\
\arrayrulecolor{gray!60}\cmidrule[0.4pt]{1-15}\arrayrulecolor{black}
\multirow{4}{*}{\shortstack{\textbf{T2I}\\\textbf{(Multi/length)}}}
& CMCL   & TKDE'24  & \underline{79.09} & \textbf{80.28} & 80.50 & 80.45 & \underline{58.72} & 64.46 & \textbf{69.52} & \textbf{72.80} & \textbf{69.95} & \textbf{72.87} & \underline{75.30} & 75.73 \\
& RCMH   & TPAMI'26 & 74.11 & 74.09 & 74.21 & 74.25 & 54.69 & 57.29 & 58.71 & 59.41 & 67.03 & 70.52 & 73.02 & 75.31 \\
& VH-ABL & TMM'26   & 74.96 & 75.94 & 77.50 & 78.58 & 57.22 & 61.28 & 65.60 & 67.01 & 63.72 & 70.11 & 73.55 & 75.27 \\
\rowcolor{multibitrow}
& \textbf{MultiBit} & \textbf{OURS} & \textbf{79.25} & \underline{80.02} & \textbf{80.82} & \textbf{81.25} & \textbf{59.94} & \textbf{65.59} & \underline{69.45} & \underline{71.86} & \underline{68.81} & \underline{71.83} & \textbf{76.15} & \textbf{77.84} \\
\bottomrule
\end{tabular}%
}
\vspace{2pt}

\end{table}

\noindent\textbf{Overall Results.} 
Table~\ref{tab:sota-comparison} reports 24 settings, spanning two retrieval directions, three datasets and four code lengths. MultiBit attains the best score in 16 of them, including every 64- and 128-bit cell on MIRFlickr-25K and MS-COCO and all four lengths on IAPR TC-12 in the image-to-text direction. CMCL takes the remaining eight — six at 16 or 32 bits and two in the text-to-image direction of IAPR TC-12 — and RCMH and VH-ABL remain below MultiBit throughout.

\begin{figure}[htbp]
\centering
\includegraphics[width=\textwidth]{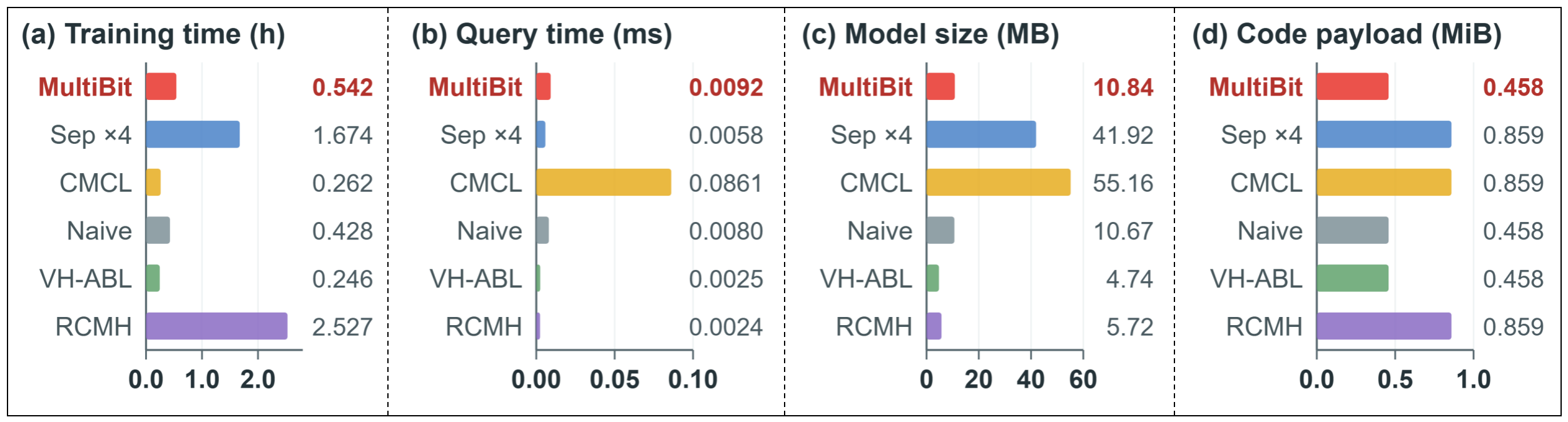}
\caption{Deployment cost of serving all four code lengths with one model on MIRFlickr-25K: (a) total training time for the four lengths; (b) 64-bit query time excluding the CLIP forward pass; (c) trainable model size excluding the shared frozen backbone; (d) packed code payload. }
\label{fig:performance-efficiency}
\end{figure}

\noindent\textbf{Performance--Efficiency.}\enspace Figure~\ref{fig:performance-efficiency} reports the deployment cost on MIRFlickr-25K, where one model serves all four code lengths: the 64-bit query time is 0.0092 ms, about $9\times$ lower than the accuracy-competitive CMCL (0.0861 ms); Table~\ref{tab:multi-length-cost} covers all three datasets and Appendix Table~\ref{tab:performance-efficiency-raw} the fixed-length methods.

\subsection{Ablation Study}
\label{sec:ablation-analysis}

\begin{table}[htbp]
\centering
\caption{Progressive ablation of the core components on three datasets and four code lengths.}
\label{tab:component-ablation}
\scriptsize
\setlength{\tabcolsep}{2.6pt}
\renewcommand{\arraystretch}{0.88}
\resizebox{\textwidth}{!}{%
\begin{tabular}{lcccccccccccc}
\toprule
\textbf{Variant}
& \multicolumn{4}{c}{\textbf{MIRFlickr-25K}}
& \multicolumn{4}{c}{\textbf{IAPR TC-12}}
& \multicolumn{4}{c}{\textbf{MS-COCO}} \\
\cmidrule(lr){2-5}\cmidrule(lr){6-9}\cmidrule(lr){10-13}
& \textbf{16} & \textbf{32} & \textbf{64} & \textbf{128}
& \textbf{16} & \textbf{32} & \textbf{64} & \textbf{128}
& \textbf{16} & \textbf{32} & \textbf{64} & \textbf{128} \\
\midrule
Baseline (no FRT/SBO/MPA)
& 80.500 & 81.488 & 81.819 & 81.861
& 60.427 & 65.008 & 67.206 & 68.555
& 66.733 & 70.423 & 72.274 & 72.925 \\
+ FRT
& 80.803 & 81.845 & 82.540 & 82.749
& 60.433 & 65.211 & 67.939 & 69.642
& 68.008 & 71.311 & 73.886 & 74.997 \\
+ FRT + SBO
& 81.013 & 82.122 & 83.005 & 83.374
& 60.434 & 65.391 & 68.592 & 70.733
& 68.635 & 71.885 & 75.106 & 76.834 \\
\rowcolor{multibitrow}
\textbf{Full MultiBit}
& \textbf{81.057} & \textbf{82.202} & \textbf{83.162} & \textbf{83.592}
& \textbf{60.449} & \textbf{65.426} & \textbf{68.846} & \textbf{70.965}
& \textbf{69.178} & \textbf{72.177} & \textbf{75.921} & \textbf{77.372} \\
\bottomrule
\end{tabular}%
}
\end{table}

\begin{figure}[htbp]
\centering
\includegraphics[width=\textwidth]{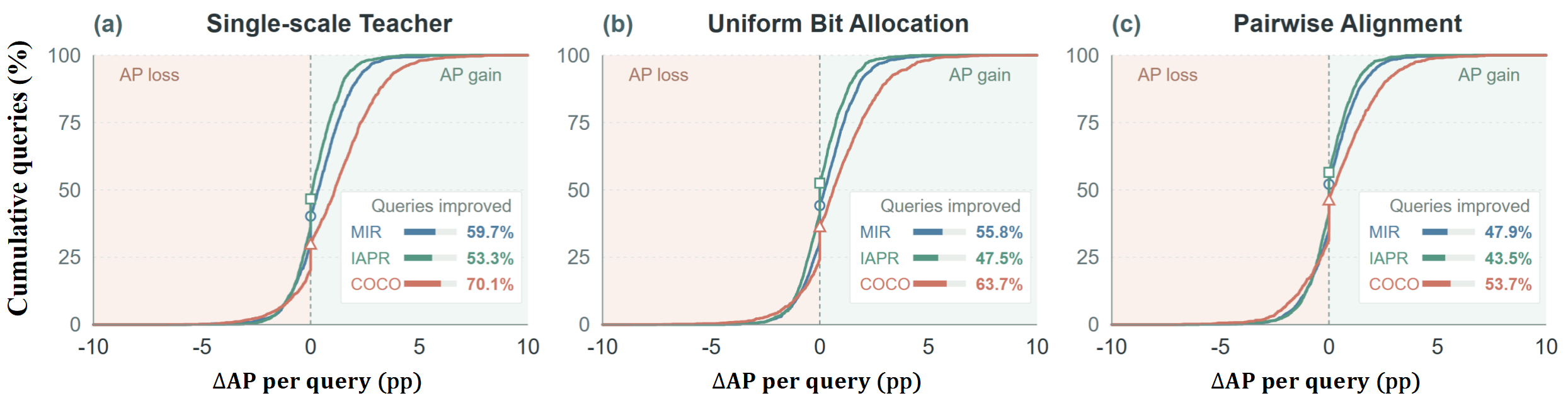}
\caption{Query-level comparison of Full MultiBit with the three alternative designs on the 64-bit code, over 1,500 queries per dataset: cumulative distributions of the per-query AP difference $\Delta$AP, with the improved share in the legend.}
\label{fig:query-level-gain}
\end{figure}

\noindent\textbf{Component Ablation.} As shown in Table~\ref{tab:component-ablation}, each component improves Mean mAP in all twelve dataset--length settings. At 128 bits, FRT adds 0.888, 1.087, and 2.072 pp on MIRFlickr-25K, IAPR TC-12, and MS-COCO, SBO a further 0.625, 1.091, and 1.837 pp, and MPA less; the last 64 bits, which span scales 1--4, gain 0.43/2.12/1.45 pp with the full model against 0.04/1.35/0.65 for the baseline. Query-level gains are positive against every design: 43.5\% to 70.1\% of the 1,500 queries in each group improve, with a mean gain of 0.03 to 1.18 pp (Figure~\ref{fig:query-level-gain}; Appendix Table~\ref{tab:query-level-gain}).

\subsection{Relevance Resolution Analysis}
\label{sec:resolution-analysis}

\begin{table}[htbp]
\centering
\caption{Relevance resolution of MultiBit and recent methods on three datasets at 16, 32, and 64 bits, under the IDF-weighted evaluator.}
\label{tab:resolution-sota}
\tiny
\setlength{\tabcolsep}{1.5pt}
\renewcommand{\arraystretch}{0.95}
\resizebox{\textwidth}{!}{%
\begin{tabular}{llcccccccccccc}
\toprule
\textbf{Dataset} & \textbf{Method}
& \multicolumn{4}{c}{\textbf{16 bits}}
& \multicolumn{4}{c}{\textbf{32 bits}}
& \multicolumn{4}{c}{\textbf{64 bits}} \\
\cmidrule(lr){3-6}\cmidrule(lr){7-10}\cmidrule(lr){11-14}
& & \textbf{N$\uparrow$} & \textbf{$\tau_b\uparrow$} & \textbf{C$\downarrow$} & \textbf{I$\downarrow$}
& \textbf{N$\uparrow$} & \textbf{$\tau_b\uparrow$} & \textbf{C$\downarrow$} & \textbf{I$\downarrow$}
& \textbf{N$\uparrow$} & \textbf{$\tau_b\uparrow$} & \textbf{C$\downarrow$} & \textbf{I$\downarrow$} \\
\midrule
\multirow{5}{*}{MIRFlickr-25K}
& DECH     & 0.632 & 0.503 & 0.396 & 0.254 & 0.650 & 0.540 & 0.374 & 0.237 & 0.666 & 0.557 & 0.356 & 0.220 \\
& DGHDGH   & 0.624 & 0.489 & 0.405 & 0.260 & 0.644 & 0.527 & 0.382 & 0.243 & 0.659 & 0.548 & 0.363 & 0.228 \\
& CMCL     & 0.643 & 0.518 & 0.387 & 0.246 & 0.662 & 0.554 & 0.361 & 0.228 & 0.677 & 0.568 & 0.343 & 0.212 \\
& VH-ABL   & 0.651 & 0.531 & 0.378 & 0.241 & 0.670 & 0.564 & 0.352 & 0.220 & 0.686 & 0.579 & 0.333 & 0.204 \\
\rowcolor{multibitrow}
& \textbf{MultiBit} & \textbf{0.660} & \textbf{0.555} & \textbf{0.366} & \textbf{0.232} & \textbf{0.678} & \textbf{0.573} & \textbf{0.342} & \textbf{0.214} & \textbf{0.697} & \textbf{0.590} & \textbf{0.321} & \textbf{0.195} \\
\midrule
\multirow{5}{*}{IAPR TC-12}
& DECH     & 0.611 & 0.424 & 0.420 & 0.311 & 0.634 & 0.451 & 0.397 & 0.291 & 0.653 & 0.468 & 0.375 & 0.273 \\
& DGHDGH   & 0.604 & 0.408 & 0.429 & 0.317 & 0.627 & 0.447 & 0.404 & 0.295 & 0.646 & 0.463 & 0.382 & 0.279 \\
& CMCL     & 0.620 & 0.443 & 0.408 & 0.302 & 0.647 & 0.460 & 0.383 & 0.281 & 0.663 & 0.478 & 0.365 & 0.263 \\
& VH-ABL   & 0.629 & 0.450 & 0.397 & 0.295 & 0.656 & 0.472 & 0.373 & 0.273 & 0.676 & 0.489 & 0.350 & 0.253 \\
\rowcolor{multibitrow}
& \textbf{MultiBit} & \textbf{0.640} & \textbf{0.461} & \textbf{0.385} & \textbf{0.285} & \textbf{0.668} & \textbf{0.481} & \textbf{0.361} & \textbf{0.264} & \textbf{0.688} & \textbf{0.499} & \textbf{0.339} & \textbf{0.244} \\
\midrule
\multirow{5}{*}{MS-COCO}
& DECH     & 0.672 & 0.497 & 0.372 & 0.263 & 0.690 & 0.515 & 0.349 & 0.244 & 0.707 & 0.531 & 0.330 & 0.226 \\
& DGHDGH   & 0.665 & 0.492 & 0.380 & 0.269 & 0.683 & 0.509 & 0.356 & 0.249 & 0.700 & 0.524 & 0.336 & 0.232 \\
& CMCL     & 0.684 & 0.508 & 0.360 & 0.253 & 0.704 & 0.526 & 0.336 & 0.233 & 0.719 & 0.542 & 0.316 & 0.216 \\
& VH-ABL   & 0.692 & 0.516 & 0.351 & 0.247 & 0.712 & 0.537 & 0.326 & 0.226 & 0.729 & 0.553 & 0.307 & 0.209 \\
\rowcolor{multibitrow}
& \textbf{MultiBit} & \textbf{0.700} & \textbf{0.528} & \textbf{0.340} & \textbf{0.237} & \textbf{0.721} & \textbf{0.547} & \textbf{0.316} & \textbf{0.216} & \textbf{0.739} & \textbf{0.564} & \textbf{0.297} & \textbf{0.199} \\

\bottomrule
\end{tabular}%
}
\end{table}

\noindent\textbf{Relevance Resolution.} Table~\ref{tab:resolution-sota} extends Table~\ref{tab:sota-comparison} with CMCL and VH-ABL; Appendix Table~\ref{tab:resolution-multilength} adds RCMH and Naive Prefix. MultiBit achieves the highest NDCG@100~\citep{jarvelin2002cumulated} and Kendall's $\tau_b$~\citep{kendall1938new} and the lowest collision and inversion rates, indicating better preservation of the label-derived graded order in finite Hamming spaces. Because this evaluator shares its prior with the relation teacher, we repeat the comparison under a frequency-free label-name reference, where MultiBit again leads on every metric in all nine settings and FRT remains the strongest teacher (Appendix Sections~\ref{sec:resolution-evaluation-details} and~\ref{sec:appendix-decoupled-evaluator}; Tables~\ref{tab:teacher-quality} and~\ref{tab:resolution-semantic}).

\begin{figure}[htbp]
\centering
\includegraphics[width=\textwidth]{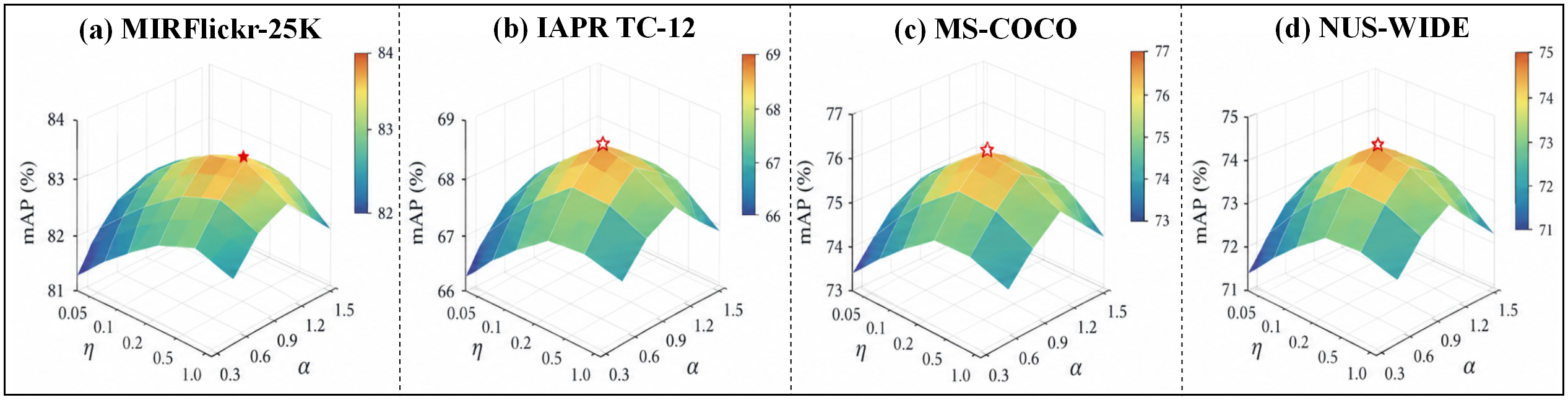}
\caption{Validation Mean mAP (\%) under different combinations of the FRT hyperparameters $\alpha$ and $\eta$ at 64 bits on all datasets. Red stars mark the best configuration for each dataset, and the response surfaces form a broad high-performance region around intermediate values of both parameters.}
\label{fig:parameter-sensitivity}
\end{figure}

\noindent\textbf{Parameter Sensitivity.} Figure~\ref{fig:parameter-sensitivity} shows a broad high-performance region around intermediate values of $\alpha$ and $\eta$ on every dataset, whereas extreme settings degrade moderately; FRT therefore does not depend on a narrow configuration (complete grid in Appendix Table~\ref{tab:parameter-sensitivity-grid}).

\subsection{Multi-Length Consistency and Efficiency}
\label{sec:multi-length-analysis}
\begin{table}[htbp]
\centering
\caption{Retrieval accuracy and deployment cost of serving four code lengths on three datasets (Mean mAP in \%). Models gives the number of trained models needed for all four lengths, Model MB the trainable-component size excluding the frozen CLIP backbone, Code MiB the packed code payload, Train (h) the cost of all four lengths, and Query (ms) the 64-bit query time excluding the CLIP forward pass.}
\label{tab:multi-length-cost}
\scriptsize
\setlength{\tabcolsep}{3pt}
\renewcommand{\arraystretch}{0.84}
\resizebox{\textwidth}{!}{%
\begin{tabular}{llccccccccc}
\toprule
\textbf{Dataset} & \textbf{Method}
& \multicolumn{4}{c}{\textbf{Mean mAP (\%)}}
& \textbf{Models}
& \textbf{Model MB}
& \textbf{Code MiB}
& \textbf{Train (h)}
& \textbf{Query (ms)} \\
\cmidrule(lr){3-6}
& & \textbf{16} & \textbf{32} & \textbf{64} & \textbf{128} & & & & & \\
\midrule
\multirow{6}{*}{MIRFlickr-25K}
& Naive Prefix & 78.765 & 80.922 & 82.311 & 83.284 & 1 & 10.67 & \textbf{0.458} & 0.428 & 0.0080 \\
& CMCL   & \textbf{81.107} & \textbf{82.312} & 82.856 & 82.946 & 1 & 55.16 & 0.859 & 0.262 & 0.0861 \\
& RCMH   & 72.457 & 72.851 & 73.152 & 73.348 & 1 & 5.72 & 0.859 & 2.527 & 0.0024 \\
& VH-ABL & 77.621 & 78.835 & 80.001 & 80.657 & 1 & \textbf{4.74} & \textbf{0.458} & 0.246 & 0.0025 \\
\rowcolor{multibitrow}
& MultiBit-Sep & 80.921 & 82.104 & 83.090 & 83.478 & 4 & 41.92 & 0.859 & 1.674 & 0.0058 \\
\rowcolor{multibitrow}
& \textbf{MultiBit} & 81.057 & 82.202 & \textbf{83.162} & \textbf{83.592} & 1 & 10.84 & \textbf{0.458} & 0.542 & 0.0092 \\
\midrule
\multirow{6}{*}{IAPR TC-12}
& Naive Prefix & 57.884 & 62.744 & 66.420 & 70.422 & 1 & 10.67 & \textbf{0.446} & 0.451 & 0.0069 \\
& CMCL   & 58.374 & 63.577 & 68.216 & \textbf{71.198} & 1 & 55.16 & 0.837 & 0.269 & 0.0855 \\
& RCMH   & 55.797 & 58.386 & 59.635 & 60.231 & 1 & 5.72 & 0.837 & 2.690 & 0.0022 \\
& VH-ABL & 55.964 & 60.790 & 64.601 & 66.192 & 1 & \textbf{4.74} & \textbf{0.446} & 0.249 & 0.0026 \\
\rowcolor{multibitrow}
& MultiBit-Sep & 60.301 & 65.291 & 68.701 & 70.861 & 4 & 41.92 & 0.837 & 1.781 & 0.0058 \\
\rowcolor{multibitrow}
& \textbf{MultiBit} & \textbf{60.449} & \textbf{65.426} & \textbf{68.846} & 70.965 & 1 & 10.84 & \textbf{0.446} & 0.579 & 0.0076 \\
\midrule
\multirow{6}{*}{MS-COCO}
& Naive Prefix & 64.201 & 68.693 & 73.808 & 76.955 & 1 & 10.67 & \textbf{3.577} & 0.463 & 0.0088 \\
& CMCL   & \textbf{70.090} & \textbf{72.908} & 75.215 & 75.702 & 1 & 55.16 & 6.707 & 0.802 & 0.0901 \\
& RCMH   & 67.427 & 70.653 & 73.041 & 75.157 & 1 & 5.72 & 6.707 & 2.602 & 0.0053 \\
& VH-ABL & 63.523 & 70.228 & 73.583 & 75.322 & 1 & \textbf{4.74} & \textbf{3.577} & 0.213 & 0.0058 \\
\rowcolor{multibitrow}
& MultiBit-Sep & 68.960 & 71.994 & 75.764 & 77.284 & 4 & 41.92 & 6.707 & 1.834 & 0.0089 \\
\rowcolor{multibitrow}
& \textbf{MultiBit} & 69.178 & 72.177 & \textbf{75.921} & \textbf{77.372} & 1 & 10.84 & \textbf{3.577} & 0.601 & 0.0096 \\
\bottomrule
\end{tabular}}
\end{table}

\noindent\textbf{Accuracy and Deployment Cost.}\enspace Table~\ref{tab:multi-length-cost} reports the full comparison. Averaged over the four code lengths, MultiBit attains the highest Mean mAP on all three datasets (82.50, 66.42, and 73.66) and stays above the four separately trained models of MultiBit-Sep at all twelve settings. Across individual lengths it is competitive or better, with CMCL ahead in five of the twelve cells by at most 0.92 pp and MultiBit leading in the other seven. One nested model therefore replaces four at roughly 74\% and 47\% less model and code storage.

\begin{figure}[htbp]
\centering
\includegraphics[width=\textwidth]{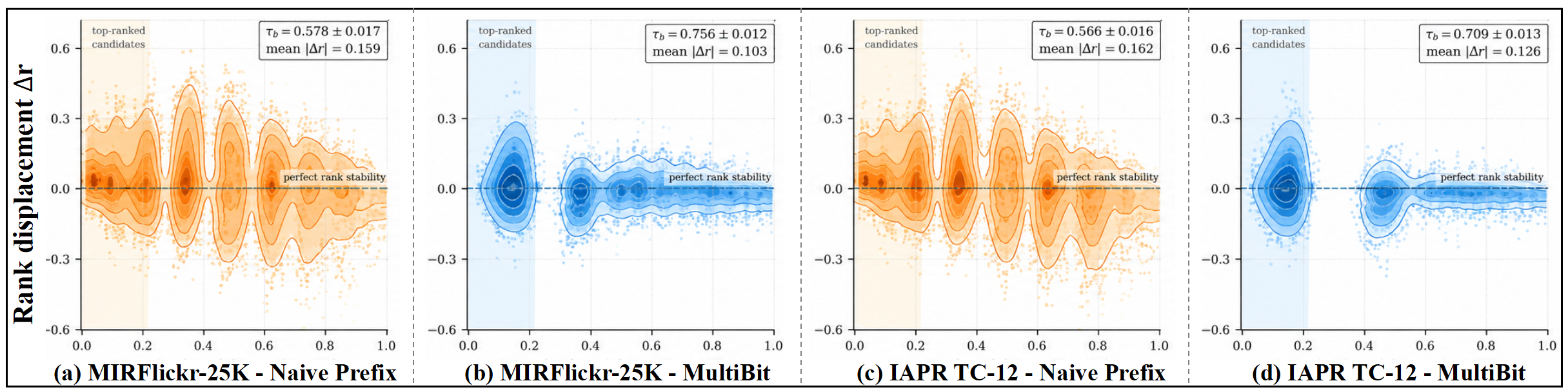}
\caption{I2T rank shifts of Naive Prefix and MultiBit between 16- and 128-bit codes on two datasets, with five sampled queries per panel and annotated tie-corrected $\tau_b$ and mean $|\Delta r|$.
}

\label{fig:cross-length-stability}
\end{figure}

\noindent\textbf{Cross-Length Ranking Consistency.}\enspace Figure~\ref{fig:cross-length-stability} visualizes I2T rank changes from 16 to 128 bits. Against Naive Prefix, the displacements of MultiBit concentrate closer to zero (mean $|\Delta r|$ 0.103 versus 0.159 and 0.126 versus 0.162; Appendix Table~\ref{tab:rank-displacement-sample}), with higher tie-corrected $\tau_b$. The same holds against every published multi-length baseline, including CMCL (0.760 versus 0.633 at 16--128 bits on MIRFlickr-25K); removing MPA lowers the six-pair I2T mean by 0.006 to 0.035. Complete results are in Appendix Table~\ref{tab:cross-length-agreement-complete}.

\subsection{Qualitative Retrieval Analysis}
\label{sec:qualitative-retrieval}

\begin{figure}[htbp]
\centering
\includegraphics[width=0.94\textwidth]{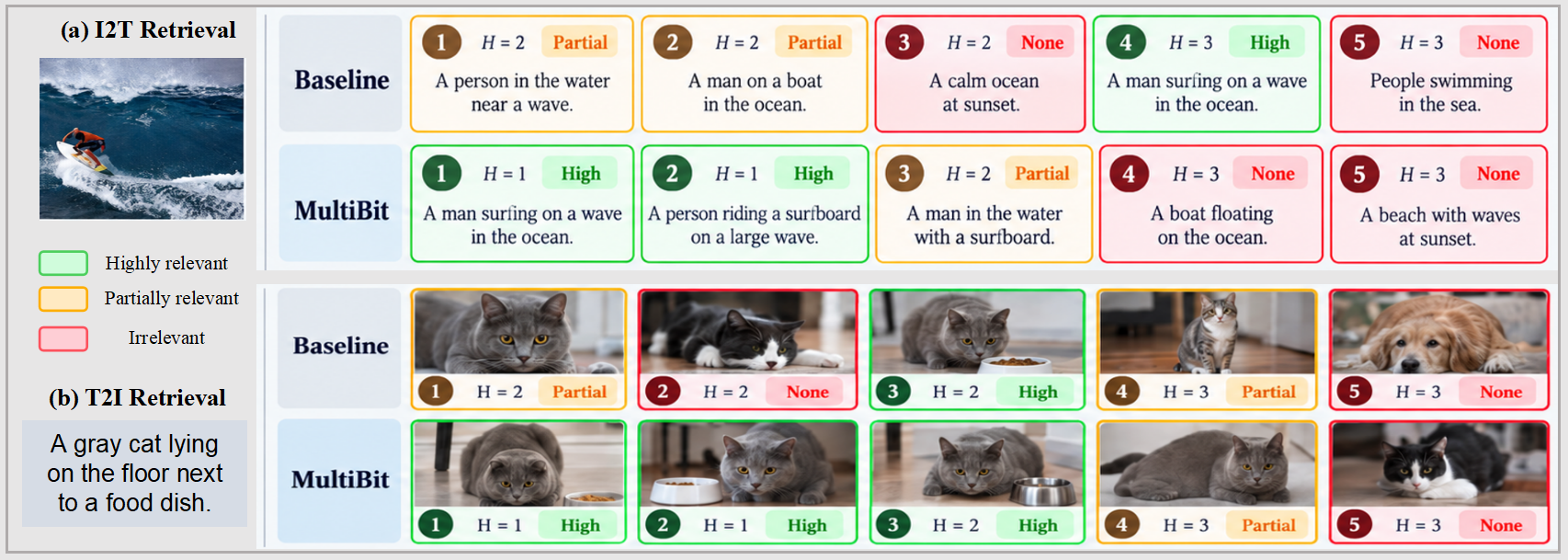}
\caption{Illustrative 16-bit I2T and T2I rankings on MIRFlickr-25K, annotated with rank, distance, and relevance; MultiBit brings the highly relevant candidates closer than the baseline.}
\label{fig:qualitative-retrieval}
\end{figure}

\noindent\textbf{Qualitative Retrieval Analysis.}\enspace Figure~\ref{fig:qualitative-retrieval} shows 16-bit examples. In the image-to-text case the baseline ranks a highly relevant caption at $H=3$ behind three weaker candidates at $H=2$ (an inversion), and in the text-to-image case a highly relevant image shares $H=2$ with an irrelevant one that precedes it (a collision). 
Equal-distance candidates are ordered by a fixed rule rather than by relevance, so the gain is distance separation, not re-ranking; the examples do not estimate prevalence.

\section{Conclusion}
\label{sec:conclusion}

We introduced the relevance resolution bottleneck and proposed \method, which transfers relevance resolution to multi-length Hamming spaces through a scale-decomposable fractional relation teacher, scale-aware bit organization, and multi-length prefix alignment. A single nested model attains the highest Mean mAP averaged over the four code lengths on all three datasets, and attains the highest cross-length ranking agreement among all compared methods. These results indicate that organizing relations by propagation scale is a practical route to consistent multi-length binary retrieval.

\subsection*{AI Use Statement}

In this work, generative AI tools were only used to edit the manuscript for improved readability and linguistic quality. Generative AI was not used for data generation, theoretical model development, formulation or proof of mathematical claims, method design and implementation, or result interpretation; none of these disclosure-required tasks apply to this work. All AI-suggested edits were reviewed against the original text by the authors, who take full responsibility for the final manuscript content, including all text, claims, and research artifacts.

\bibliography{MultiBit-references}
\bibliographystyle{iclr2027/iclr2027_conference}

\clearpage
\appendix
\renewcommand{\thesubsection}{\thesection.\arabic{subsection}}
\numberwithin{table}{section}
\renewcommand{\thetable}{\thesection-\arabic{table}}
\numberwithin{equation}{section}
\renewcommand{\theequation}{\thesection.\arabic{equation}}
\begin{center}
{\LARGE\bfseries Appendix}\\[6pt]
{\large Relevance-Resolution Transfer via Scale-Decomposable Fractional Diffusion for Multi-Length Cross-Modal Hash Retrieval}
\end{center}
\vspace{0.5em}
\section{\MBLang{Method Notation and Experimental Settings}{方法符号与实验设置}}
\label{sec:appendix-parameters}
\subsection{\MBLang{Data Split, Evaluation Metrics, and Implementation Details}{数据划分、评估指标与实现细节}}
\label{sec:appendix-setup}

\noindent\textbf{\MBLang{Data Split.}{数据划分。}}\enspace\MBLang{All methods share the same split, drawn with the fixed random seed $42$, so that the training, validation, query, and database indices are identical across methods and the comparison is fair. Table~\ref{tab:data-split} summarizes the resulting partitions. The retrieval database is the full dataset minus the query set; it therefore contains the $10{,}000$ training and validation pairs whose labels are used for supervision, and Table~\ref{tab:data-split} lists the resulting database sizes. The $1{,}000$ validation pairs are disjoint from the $5{,}000$ queries and are used only to select the checkpoint, never as reported results.}{所有方法共用同一划分，均以固定随机种子 $42$ 抽取，从而保证各方法的训练、验证、查询与检索库索引完全一致，比较是公平的。表~\ref{tab:data-split} 汇总了各数据集的划分结果。检索库为全集除去查询集，因此包含那 $10{,}000$ 个标签被用于监督的训练与验证样本，表~\ref{tab:data-split} 给出由此得到的检索库规模；$1{,}000$ 个验证样本与 $5{,}000$ 个查询不相交，仅用于选择 checkpoint，不作为结果报告。}

\begin{table}[!htbp]
\centering
\caption{\MBLang{Data split shared by all compared methods.}{所有对比方法共用的数据划分。}}
\label{tab:data-split}
\small
\setlength{\tabcolsep}{5.5pt}
\begin{tabular}{lrrrrrr}
\toprule
\textbf{\MBLang{Dataset}{数据集}} & \textbf{\MBLang{Pairs}{样本对数}} & \textbf{\MBLang{Classes}{类别数}} & \textbf{\MBLang{Train}{训练}} & \textbf{\MBLang{Validation}{验证}} & \textbf{\MBLang{Query}{查询}} & \textbf{\MBLang{Database}{检索库}} \\
\midrule
MIRFlickr-25K & 20,015 & 24 & 9,000 & 1,000 & 5,000 & 15,015 \\
NUS-WIDE      & 195,834 & 21 & 9,000 & 1,000 & 5,000 & 190,834 \\
MS-COCO       & 122,218 & 80 & 9,000 & 1,000 & 5,000 & 117,218 \\
IAPR TC-12    & 19,627 & 291 & 9,000 & 1,000 & 5,000 & 14,627 \\
\bottomrule
\end{tabular}
\end{table}

\noindent\textbf{\MBLang{Evaluation Metrics.}{评估指标。}}\enspace\MBLang{We report Image-to-Text (I2T) and Text-to-Image (T2I) retrieval and their arithmetic mean (Mean). The primary metric is mAP@all: for each query, average precision is computed over the complete retrieval database with $k$ set to the database size, and the per-query values are then averaged over the whole query set. The resolution analysis additionally uses graded relevance, NDCG@100, tie-corrected Kendall's $\tau_b$, and the collision and inversion rates defined in Section~\ref{sec:resolution-evaluation-details}; the multi-length analysis uses the same $\tau_b$ together with the rank-displacement measure defined in Section~\ref{sec:appendix-multilength}.}{我们报告图像到文本（I2T）与文本到图像（T2I）检索及其算术平均（Mean）。主指标为 mAP@all：对每个查询在完整检索库上以库规模作为 $k$ 计算平均精度，再对全部查询取平均。分辨率分析还使用分级相关性、NDCG@100、经并列校正的 Kendall's $\tau_b$ 以及第~\ref{sec:resolution-evaluation-details} 节定义的碰撞率与反转率；多码长分析使用同一 $\tau_b$ 及第~\ref{sec:appendix-multilength} 节定义的排序位移。}

\noindent\textbf{\MBLang{Implementation Details.}{实现细节。}}\enspace\MBLang{All experiments are implemented in PyTorch and run on a single NVIDIA GeForce RTX 4090 GPU. All methods adopt CLIP ViT-B/32 as the shared backbone. Code lengths are $16$, $32$, $64$, and $128$ bits. Models are trained with Adam for $100$ epochs at a batch size of $128$, with an initial learning rate of $0.001$ and a cosine annealing schedule, and the checkpoint with the best validation Mean mAP is selected. The FRT hyperparameters $\alpha$ and $\eta$ are selected on the validation set. The final evaluation is performed on the complete retrieval database. Listwise teacher and student distributions are normalized over the current mini-batch, and no memory bank is used. All compared baselines are our own reproductions retrained under this shared protocol. The ablation baseline in Table~\ref{tab:component-ablation} shares the backbone, encoder, data split, and training schedule of MultiBit, but is trained with the standard cross-modal hashing objective on the CLIP features only: it uses no relation teacher, no scale-aware bit organization, and neither the scale objective nor the prefix objective, and its results at 16, 32, and 64 bits are obtained by training a single 128-bit model and truncating it with the same prefix operator $P_B$.}{全部实验基于 PyTorch 实现，运行于单张 NVIDIA GeForce RTX 4090 GPU。所有方法均采用 CLIP ViT-B/32 作为共享骨干。码长设置为 $16$、$32$、$64$ 与 $128$ bit。模型使用 Adam 训练 $100$ 轮，批大小为 $128$，初始学习率 $0.001$ 并采用余弦退火调度，选取验证 Mean mAP 最优的 checkpoint。FRT 超参数 $\alpha$ 与 $\eta$ 在验证集上选取。最终评测在完整检索库上进行。列表目标中的教师与学生分布均在同一 mini-batch 上归一化，且全部实验均未使用 memory bank。所有对比基线均为我们在该统一协议下重新复现并训练的结果。表~\ref{tab:component-ablation} 的消融 baseline 与 MultiBit 共用骨干、编码器、数据划分与训练设置，但仅使用 CLIP 特征上的标准跨模态哈希目标训练：不使用关系教师、不使用尺度感知 bit 组织，也不使用尺度目标与前缀目标；其 16、32、64 bit 结果由训练单个 128-bit 模型并按同一前缀算子 $P_B$ 截断得到。}

\noindent\textbf{\MBLang{Backbone, Heads, and Per-Baseline Settings.}{骨干、哈希头与各基线设置。}}\enspace\MBLang{The CLIP ViT-B/32 backbone is frozen and only the projection and the scale hash heads are trained. Each of the $M=4$ scale heads is an independent linear map from the shared $512$-dimensional feature to its own $m_\ell$ dimensions, with no hidden layer and no parameters shared across heads. All four code lengths are read from the same model: one checkpoint is selected on validation and every length is the prefix of its maximum-length code, and the validation criterion is the mean over the four target lengths of the validation Mean mAP. The compared baselines are trained with their own recommended optimizers and learning rates rather than one shared setting (Table~\ref{tab:baseline-optimizers}); only the data split, the number of epochs, the batch size and the evaluation protocol are shared.}{CLIP ViT-B/32 骨干被冻结，仅训练投影层与各尺度哈希头。$M=4$ 个尺度头各自是一个从共享 $512$ 维特征到自身 $m_\ell$ 维的独立线性映射，没有隐藏层，头与头之间不共享参数。四个码长由同一个模型读出：只按验证集选出唯一 checkpoint，每个码长取其最大长度码的前缀，验证准则为四个目标码长上验证 Mean mAP 的平均。对比基线各自采用其推荐的优化器与学习率，而非统一设置（表~\ref{tab:baseline-optimizers}）；各方法仅共享数据划分、训练轮数、批大小与评测流程。}

\begin{table}[!htbp]
\centering
\caption{\MBLang{Per-baseline optimizers and learning rates, taken from the original implementations. The data split, the number of epochs, the batch size and the evaluation protocol are shared across methods.}{各基线的优化器与学习率，取自各自原始实现。数据划分、训练轮数、批大小与评测流程在各方法间统一。}}
\label{tab:baseline-optimizers}
\small
\setlength{\tabcolsep}{6pt}
\resizebox{\textwidth}{!}{%
\begin{tabular}{lll}
\toprule
\textbf{\MBLang{Method}{方法}} & \textbf{\MBLang{Optimizer}{优化器}} & \textbf{\MBLang{Learning rate}{学习率}} \\
\midrule
DECH, DGHDGH, DIMCH, DPBE, DPSIH, DSTH, RCMH & AdamW & $1\times10^{-4}$ \\
DNPH & BertAdam & $1\times10^{-3}$ \\
CMCL & BertAdam & $0.001$ (MIRFlickr-25K) / $0.002$ (MS-COCO, NUS-WIDE) \\
DSSH & Adam & $0.1$; CLIP learning rate $0.001$ \\
VH-ABL & AdamW & $1\times10^{-4}$ \\
\bottomrule
\end{tabular}
}
\end{table}

\subsection{\MBLang{Complete Parameter Register}{全文参数总表}}

\begin{table}[!htbp]
\centering
\caption{\MBLang{Notation, model controls, and experimental settings used throughout the manuscript.}{全文符号、模型控制参数及实验设置。}}
\label{tab:all-parameters}
\small
\setlength{\tabcolsep}{4.5pt}
\renewcommand{\arraystretch}{1.06}
\begin{tabular}{@{}l l p{0.23\textwidth} p{0.40\textwidth}@{}}
\toprule
\textbf{\MBLang{Group}{类别}} & \textbf{\MBLang{Symbol / setting}{符号／设置}} & \textbf{\MBLang{Definition}{定义}} & \textbf{\MBLang{Value or setting}{取值或设置}} \\
\midrule
\MBLang{Data}{数据} & $N,C$ & \MBLang{Training-pair count; label count}{训练样本对数；标签数} & \MBLang{See Table~\ref{tab:data-split}}{见表~\ref{tab:data-split}} \\
\MBLang{Data}{数据} & $\mathcal B,K,B_{\max}$ & \MBLang{Target lengths; number of lengths; maximum length}{目标码长集合；码长数；最大码长} & $\{16,32,64,128\}$, $4$, $128$ \\
\MBLang{Data}{数据} & $d$ & \MBLang{Shared feature dimension before hash heads}{哈希头之前的共享特征维度} & $512$ \\
\MBLang{Data}{数据} & $P_B$ & \MBLang{First-$B$-bit prefix selection}{前 $B$ 位的前缀选择矩阵} & \MBLang{Determined by $B$ and $B_{\max}$}{由 $B$ 与 $B_{\max}$ 确定} \\
\midrule
\MBLang{FRT}{FRT} & $\varepsilon$ & \MBLang{PPMI smoothing constant}{PPMI 平滑常数} & $0.1$ \\
\MBLang{FRT}{FRT} & $W,D,L$ & \MBLang{PPMI label graph, degree matrix, normalized Laplacian}{PPMI 标签图、度矩阵与归一化拉普拉斯矩阵} & \MBLang{Derived from training labels}{由训练标签推得} \\
\MBLang{FRT}{FRT} & $\alpha$ & \MBLang{Fractional diffusion order}{分数阶扩散阶数} & \MBLang{$1.20$ (MIRFlickr-25K); $0.90$ (others)}{$1.20$（MIRFlickr-25K）；$0.90$（其余数据集）} \\
\MBLang{FRT}{FRT} & $\eta$ & \MBLang{Positive kernel shift and diffusion-range control}{正定核平移及扩散范围控制} & $0.20$ \\
\MBLang{FRT}{FRT} & $M$ & \MBLang{Gauss--Laguerre quadrature points / retained scales}{Gauss--Laguerre 求积点数／保留尺度数} & $4$ \\
\MBLang{FRT}{FRT} & $\tau_\ell,\rho_\ell$ & \MBLang{Diffusion times and quadrature weights, local to long range}{由局部到远程的扩散时间与求积权重} & \MBLang{$\alpha{=}1.20$: $\tau=(2.00,9.55,23.89,48.56)$, $\rho=(3.88,2.70,0.32,0.005)$; $\alpha{=}0.90$: $\tau=(1.43,8.32,22.08,46.18)$, $\rho=(2.67,1.44,0.15,0.002)$}{$\alpha{=}1.20$：$\tau=(2.00,9.55,23.89,48.56)$，$\rho=(3.88,2.70,0.32,0.005)$；$\alpha{=}0.90$：$\tau=(1.43,8.32,22.08,46.18)$，$\rho=(2.67,1.44,0.15,0.002)$} \\
\MBLang{FRT}{FRT} & $s_a,\widetilde y_i$ & \MBLang{Label specificity and weighted label vector}{标签特异性与加权标签向量} & \MBLang{Derived from training-label frequencies}{由训练标签频率推得} \\
\MBLang{FRT}{FRT} & $K_{\alpha,\eta},g^*,g_\ell$ & \MBLang{Fractional kernel; integrated and scale teacher scores}{分数阶核；综合与分尺度教师分数} & \MBLang{Derived by Eqs.~(\ref{eq:fractional-kernel})--(\ref{eq:scale-teacher})}{由式~(\ref{eq:fractional-kernel})--(\ref{eq:scale-teacher}) 推出} \\
\midrule
\MBLang{SBO}{SBO} & $m_\ell$ & \MBLang{Integer bits assigned to scale $\ell$, local to long range}{由局部到远程分配给尺度 $\ell$ 的整数 bit 数} & \MBLang{$(71,50,6,1)$ for $\alpha{=}1.20$ and $(80,43,4,1)$ for $\alpha{=}0.90$, with $\sum_\ell m_\ell=B_{\max}$}{$\alpha{=}1.20$ 时为 $(71,50,6,1)$，$\alpha{=}0.90$ 时为 $(80,43,4,1)$，且 $\sum_\ell m_\ell=B_{\max}$} \\
\midrule
\MBLang{MPA}{MPA} & $T_t,T_s$ & \MBLang{Teacher and student Softmax temperatures}{教师与学生 Softmax 温度} & $0.2$ / $0.1$ \\
\MBLang{MPA}{MPA} & $\beta_B$ & \MBLang{Prefix-loss weight at length $B$}{码长 $B$ 的前缀损失权重} & \MBLang{$\sqrt{B}/\sum_{B'}\sqrt{B'}$; $(0.138,0.195,0.276,0.391)$ for $16/32/64/128$}{$\sqrt{B}/\sum_{B'}\sqrt{B'}$；$16/32/64/128$ 对应 $(0.138,0.195,0.276,0.391)$} \\
\MBLang{MPA}{MPA} & $\lambda_s$ & \MBLang{Scale-loss coefficient}{尺度损失系数} & $0.7$ \\
\MBLang{MPA}{MPA} & $\mathcal C_q$ & \MBLang{Shared cross-modal candidate set for listwise matching}{列表匹配共享的跨模态候选集合} & \MBLang{Current mini-batch}{当前 mini-batch} \\
\MBLang{MPA}{MPA} & $P^*,P_{\theta,B}$ & \MBLang{Teacher and prefix-student candidate distributions}{教师与前缀学生候选分布} & \MBLang{Defined in Section~\ref{sec:prefix-alignment}}{定义见第~\ref{sec:prefix-alignment} 节} \\
\midrule
\MBLang{Training}{训练} & \MBLang{Backbone / hardware}{骨干／硬件} & \MBLang{Common encoder and accelerator}{共用编码器及加速设备} & \MBLang{CLIP ViT-B/32; one RTX 4090}{CLIP ViT-B/32；单张 RTX 4090} \\
\MBLang{Training}{训练} & \MBLang{Optimizer / schedule}{优化器／调度} & \MBLang{Parameter update and learning-rate schedule}{参数更新与学习率调度} & \MBLang{Adam; initial LR 0.001; cosine annealing}{Adam；初始学习率 0.001；余弦退火} \\
\MBLang{Training}{训练} & \MBLang{Batch / epochs}{批大小／轮次} & \MBLang{Training batch and maximum epochs}{训练批大小与最大轮次} & $128$ / $100$ \\
\MBLang{Training}{训练} & \MBLang{Checkpoint}{选模} & \MBLang{Validation criterion}{验证集标准} & \MBLang{Best validation Mean mAP}{验证 Mean mAP 最优} \\
\MBLang{Evaluation}{评测} & $\mathcal R,\mathcal P_q,r_{qi}$ & \MBLang{Database, pairwise candidate pool, independent graded relevance}{数据库、成对比较候选池及独立分级相关性} & \MBLang{Defined in Section~\ref{sec:resolution-evaluation-details}}{定义见第~\ref{sec:resolution-evaluation-details} 节} \\
\bottomrule
\end{tabular}
\end{table}

\subsection{\MBLang{Formal Statement of the Relevance Resolution Bottleneck}{相关性分辨率瓶颈的形式化表述}}
\label{sec:appendix-rrb}

\noindent\MBLang{Let $\mathcal R$ be the retrieval database, $r_{qi}$ the graded relevance of candidate $i$ to query $q$, and $\pi_q^{*}$ the ranking of $\mathcal R$ by descending $r_{qi}$. At code length $B$, let $d_H^{(B)}(q,i)$ be the Hamming distance and $\pi_q^{(B)}$ the ranking by ascending $d_H^{(B)}$. We say that the code preserves relevance resolution at length $B$ if}{设 $\mathcal R$ 为检索数据库，$r_{qi}$ 为候选 $i$ 相对查询 $q$ 的分级相关性，$\pi_q^{*}$ 为按 $r_{qi}$ 降序得到的排序。在码长 $B$ 下，$d_H^{(B)}(q,i)$ 为 Hamming 距离，$\pi_q^{(B)}$ 为按 $d_H^{(B)}$ 升序得到的排序。若对任意 $i,j\in\mathcal R$ 有}
\begin{equation}
r_{qi}>r_{qj}
\quad\Longrightarrow\quad
d_H^{(B)}(q,i)<d_H^{(B)}(q,j),
\qquad \forall\, i,j\in\mathcal R,
\label{eq:rrb-formal}
\end{equation}
\noindent\MBLang{then the code preserves relevance resolution at length $B$, and an ideal transfer satisfies this simultaneously for every $B\in\mathcal B$. The Relevance Resolution Bottleneck (RRB) is the failure of this condition. A $B$-bit code admits only $B+1$ distinct Hamming distances, so Equation~(\ref{eq:rrb-formal}) is a capacity bound that no encoding can exceed. We therefore report collision and inversion rates as measured quantities and compare them across methods. This failure arises in three forms: (i) a \emph{collision}, where $r_{qi}>r_{qj}$ but $d_H^{(B)}(q,i)=d_H^{(B)}(q,j)$; (ii) an \emph{inversion}, where $r_{qi}>r_{qj}$ but $d_H^{(B)}(q,i)>d_H^{(B)}(q,j)$; and (iii) a \emph{cross-length inconsistency}, where the relative order of $i$ and $j$ differs between two target lengths $B$ and $B'$.}{则称该码长保持了相关性分辨率；若对所有 $B\in\mathcal B$ 同时成立，则称该传递是理想的。相关性分辨率瓶颈（RRB）即该条件不成立的情形：$B$-bit 码最多只有 $B+1$ 种不同的 Hamming 距离，因此式~(\ref{eq:rrb-formal}) 是任何编码都无法超越的容量上界；我们因而把碰撞率与反转率作为实测量报告，并据此跨方法比较。该情形表现为三种形式：（i）\emph{碰撞}，$r_{qi}>r_{qj}$ 但 $d_H^{(B)}(q,i)=d_H^{(B)}(q,j)$；（ii）\emph{反转}，$r_{qi}>r_{qj}$ 但 $d_H^{(B)}(q,i)>d_H^{(B)}(q,j)$；（iii）\emph{跨码长不一致}，候选 $i$ 与 $j$ 的相对次序在两个目标码长 $B$ 与 $B'$ 下不同。}
\section{\MBLang{Detailed Formulations and Derivations}{详细公式与推导}}
\label{sec:appendix-derivations}

\noindent\textbf{\MBLang{Label Graph Construction.}{标签图构造。}}\enspace\MBLang{Let $n_a=\sum_i y_{ia}$ and $n_{ab}=\sum_i y_{ia}y_{ib}$ denote the frequency of label $a$ and the co-occurrence frequency of labels $a$ and $b$, respectively. The PPMI label graph $W\in\mathbb R^{C\times C}$ used in Section~\ref{sec:teacher} is constructed as}{令 $n_a=\sum_i y_{ia}$ 与 $n_{ab}=\sum_i y_{ia}y_{ib}$ 分别表示标签 $a$ 的出现频次以及标签 $a$ 与 $b$ 的共现频次。第~\ref{sec:teacher} 节采用的 PPMI 标签图 $W\in\mathbb R^{C\times C}$ 构造如下：}
\begin{equation}
W_{ab}
=
\mathbb{I}[n_{ab}>0]
\max\!\left\{
0,
\log\frac{(n_{ab}+\varepsilon)N}
{(n_a+\varepsilon)(n_b+\varepsilon)}
\right\},
\quad a\ne b,
\qquad W_{aa}=0,
\label{eq:ppmi-graph}
\end{equation}
\noindent\MBLang{where $\varepsilon>0$ is the smoothing constant.}{其中 $\varepsilon>0$ 为平滑常数。}

\noindent\textbf{\MBLang{Scale-Aware Subblocks.}{尺度感知子块。}}\enspace\MBLang{Let $E^m$ be the encoder and projection for modality $m\in\{I,T\}$, mapping an input to a shared $d$-dimensional feature $z_i^m=E^m(x_i^m)$, and let $H_\ell:\mathbb R^{d}\rightarrow\mathbb R^{m_\ell}$ be the hash head of scale $\ell$. Sorting the scales from local to long range as $(\sigma_1,\ldots,\sigma_M)=\operatorname{argsort}(\tau_1,\ldots,\tau_M)$, the maximum-length representation used in Section~\ref{sec:bit-organization} is}{令 $E^m$ 表示模态 $m\in\{I,T\}$ 的编码器与投影，将输入映射为共享的 $d$ 维特征 $z_i^m=E^m(x_i^m)$；$H_\ell:\mathbb R^{d}\rightarrow\mathbb R^{m_\ell}$ 为尺度 $\ell$ 的哈希头。按照由局部到远程的尺度顺序 $(\sigma_1,\ldots,\sigma_M)=\operatorname{argsort}(\tau_1,\ldots,\tau_M)$，第~\ref{sec:bit-organization} 节采用的最大码长表示为}
\begin{equation}
\widetilde h_i^{m,(\ell)}
=
\tanh\!\left(H_\ell(z_i^m)\right)
\in(-1,1)^{m_\ell},
\qquad
\widetilde h_i^m
=
\left[
\widetilde h_i^{m,(\sigma_1)}\,\|\,\cdots\,\|\,
\widetilde h_i^{m,(\sigma_M)}
\right]
\in(-1,1)^{B_{\max}},
\label{eq:scale-subblocks}
\end{equation}

\begin{table}[!htbp]
\centering
\caption{\MBLang{Scale coverage of the four deployable code lengths. The bit budgets $m_\ell$ follow from the normalized Gauss--Laguerre weights, and the cumulative column counts the bits up to and including each scale.}{四个可部署码长的尺度覆盖。bit 预算 $m_\ell$ 由归一化的 Gauss--Laguerre 权重确定；累计列给出截至该尺度的 bit 数。}}
\label{tab:scale-coverage}
\scriptsize
\setlength{\tabcolsep}{7pt}
\renewcommand{\arraystretch}{1.04}
\begin{tabular}{cccccc}
\toprule
$\alpha$ & \MBLang{Scale $\ell$}{尺度 $\ell$} & $\tau_\ell$ & $\rho_\ell$ & $m_\ell$ & \MBLang{Cumulative}{累计} \\
\midrule
\multirow{4}{*}{1.20}
& 1 & 2.00 & 3.88 & 71 & 71 \\
& 2 & 9.55 & 2.70 & 50 & 121 \\
& 3 & 23.89 & 0.32 & 6 & 127 \\
& 4 & 48.56 & 0.005 & 1 & 128 \\
\midrule
\multirow{4}{*}{0.90}
& 1 & 1.43 & 2.67 & 80 & 80 \\
& 2 & 8.32 & 1.44 & 43 & 123 \\
& 3 & 22.08 & 0.15 & 4 & 127 \\
& 4 & 46.18 & 0.002 & 1 & 128 \\
\bottomrule
\end{tabular}
\end{table}

\noindent\MBLang{With both configurations the 16-, 32- and 64-bit prefixes lie entirely inside the first subblock; only the 128-bit code combines more than one diffusion scale. SBO therefore acts at two levels: inside the first subblock it reallocates capacity among local scale directions, which benefits every deployable length, and beyond it, it combines the remaining scales, which only the 128-bit code can do. Relative to the $+$FRT configuration of Table~\ref{tab:component-ablation}, adding SBO raises Mean mAP by $0.210/0.277/0.465$ at 16/32/64 bits and $0.625$ at 128 bits on MIRFlickr-25K, by $0.001/0.180/0.653$ and $1.091$ on IAPR TC-12, and by $0.627/0.574/1.220$ and $1.837$ on MS-COCO. The cross-scale combination is exercised at the maximum length, where the gain is largest. Table~\ref{tab:query-level-gain} provides the same-capacity control: with the total budget and the subblock structure held fixed and the bits split evenly over the four scales, Full MultiBit improves on $47.5$--$63.7\%$ of queries with a mean $\Delta$AP of $0.11$--$0.85$ pp.}{两种配置下，16、32 与 64 bit 前缀都完全落在第一个子块内；只有 128-bit 码组合了不止一个扩散尺度。因此 SBO 在两层起作用：在第一子块内部重新分配局部尺度方向的容量，这使每一个可部署码长都受益；在第一子块之外组合其余尺度，则只有 128-bit 码能够做到。相对于表~\ref{tab:component-ablation} 的 $+$FRT 配置，加入 SBO 使 Mean mAP 在 MIRFlickr-25K 上于 16/32/64 bit 提升 $0.210/0.277/0.465$、128 bit 提升 $0.625$；在 IAPR TC-12 上为 $0.001/0.180/0.653$ 与 $1.091$；在 MS-COCO 上为 $0.627/0.574/1.220$ 与 $1.837$。跨尺度组合在最大码长上被真正使用，该处增益也最大。表~\ref{tab:query-level-gain} 提供了同容量对照：在总预算与子块结构不变、仅把位数在四个尺度上平均分配时，Full MultiBit 在 $47.5\%$--$63.7\%$ 的查询上取得提升，平均 $\Delta$AP 为 $0.11$--$0.85$ 点。}

\begin{table}[!htbp]
\centering
\caption{\MBLang{Relevance resolution of MultiBit as a function of the code length, sampled at the scale boundaries of Table~\ref{tab:scale-coverage}, measured under the IDF-weighted evaluator.}{MultiBit 的相关性分辨率随码长的变化，取样于表~\ref{tab:scale-coverage} 的尺度边界，在 IDF 加权参照下测量。}}
\label{tab:scale-boundary}
\scriptsize
\setlength{\tabcolsep}{6pt}
\renewcommand{\arraystretch}{1.04}
\begin{tabular}{lcccccc}
\toprule
\textbf{\MBLang{Dataset}{数据集}} & $\alpha$ & $B$ & \textbf{N$\uparrow$} & \textbf{$\tau_b\uparrow$} & \textbf{C$\downarrow$} & \textbf{I$\downarrow$} \\
\midrule
\multirow{5}{*}{MIRFlickr-25K}
& \multirow{5}{*}{1.20} & 64 & 0.697 & 0.590 & 0.321 & 0.195 \\
& & 71 & 0.702 & 0.595 & 0.315 & 0.191 \\
& & 121 & 0.724 & 0.613 & 0.291 & 0.174 \\
& & 127 & 0.728 & 0.617 & 0.286 & 0.170 \\
& & 128 & 0.729 & 0.618 & 0.285 & 0.169 \\
\midrule
\multirow{5}{*}{IAPR TC-12}
& \multirow{5}{*}{0.90} & 64 & 0.688 & 0.499 & 0.339 & 0.244 \\
& & 80 & 0.697 & 0.506 & 0.330 & 0.237 \\
& & 123 & 0.716 & 0.522 & 0.307 & 0.220 \\
& & 127 & 0.719 & 0.525 & 0.303 & 0.217 \\
& & 128 & 0.720 & 0.526 & 0.302 & 0.216 \\
\midrule
\multirow{5}{*}{MS-COCO}
& \multirow{5}{*}{0.90} & 64 & 0.739 & 0.564 & 0.297 & 0.199 \\
& & 80 & 0.746 & 0.570 & 0.289 & 0.193 \\
& & 123 & 0.766 & 0.586 & 0.268 & 0.177 \\
& & 127 & 0.769 & 0.589 & 0.264 & 0.174 \\
& & 128 & 0.770 & 0.590 & 0.263 & 0.173 \\
\bottomrule
\end{tabular}
\end{table}

\noindent\MBLang{Resolution improves monotonically with the code length on all three datasets (Table~\ref{tab:scale-boundary}). The last 64 bits, which complete the first subblock and then cover scales 2--4, add 0.032 NDCG@100, 0.028 Kendall's $\tau_b$, $-0.036$ collision and $-0.026$ inversion on MIRFlickr-25K, with comparable changes on the other two datasets. Because this diagnostic lengthens the code and thereby increases scale coverage and bit count together, it isolates the effect of code length rather than the contribution of long-range scales alone.}{三个数据集的分辨率都随码长单调提升（表~\ref{tab:scale-boundary}）。最后 64 位补满第一个子块、继而覆盖第 2--4 尺度在 MIRFlickr-25K 上带来 +0.032 NDCG@100、+0.028 Kendall's $\tau_b$、$-0.036$ 碰撞率与 $-0.026$ 反序率，另两个数据集的变化相当。由于该诊断在加长码长的同时一并增加尺度覆盖与位数，它刻画的是码长的影响，而不是长程尺度单独的贡献。}
\noindent\MBLang{with $b_i^{m,(\ell)}=\operatorname{sign}(\widetilde h_i^{m,(\ell)})$.}{其中 $b_i^{m,(\ell)}=\operatorname{sign}(\widetilde h_i^{m,(\ell)})$。}

\noindent\textbf{\MBLang{Directional Listwise Objective.}{分方向列表目标。}}\enspace\MBLang{For a query set $\mathcal Q_m$, the directional loss of retrieval direction $m\rightarrow n$ at code length $B$ used in Section~\ref{sec:prefix-alignment} is}{对于查询集合 $\mathcal Q_m$，第~\ref{sec:prefix-alignment} 节采用的检索方向 $m\rightarrow n$ 在码长 $B$ 下的分方向损失为}
\begin{equation}
\mathcal L_B^{m\rightarrow n}
=
\frac{1}{|\mathcal Q_m|}
\sum_{q\in\mathcal Q_m}
D_{\mathrm{KL}}\!\left(
P^*(\cdot\mid q)\,\|\,P_{\theta,B}^{m\rightarrow n}(\cdot\mid q)
\right).
\label{eq:directional-list}
\end{equation}

\noindent\textbf{\MBLang{Design Rationale.}{设计依据。}}\enspace\MBLang{PPMI retains co-occurrences stronger than their independent expectation instead of letting frequent labels dominate the graph, and the specificity weight $s_a$ modulates each label by its frequency so that rare labels remain informative. In Eq.~(\ref{eq:fractional-kernel}), $e^{-tL}$ is the heat diffusion operator at scale $t$: small $t$ emphasizes local label neighborhoods, whereas large $t$ incorporates more distant dependencies, and fractional orders of this form interpolate between local and anomalous long-range transport. The cosine normalization in Eq.~(\ref{eq:integrated-teacher}) reduces the influence of label cardinality, so that directly overlapping and indirectly connected labels contribute comparably to a query-conditioned relevance score; it also means that $g^*$ is not a linear combination of the scale teachers $\{g_\ell\}$. Accordingly, $\rho_\ell$ should be read as the coefficient of $K_\ell$ in the kernel mixture rather than as a per-scale contribution to the final teacher score. Each $\tau_\ell$ in Eq.~(\ref{eq:quadrature-decomposition}) specifies a diffusion range and $\rho_\ell>0$ is its coefficient in the integrated kernel; both follow from the fractional kernel rather than manually sampled diffusion times. Because the common factor $\eta^{-\alpha}/\Gamma(\alpha)$ cancels under the normalization in Eq.~(\ref{eq:bit-budget}), the bit ratio reflects the Gauss--Laguerre weights, and hence the order $\alpha$, rather than a dataset-specific measurement of graph response; we therefore treat it as a prior over scales rather than a calibrated quantity, and evaluate it against uniform bit allocation in Table~\ref{tab:query-level-gain}. The normalized weight $\bar\rho_\ell$ in Eq.~(\ref{eq:overall-objective}) reuses the same prior to weight the per-scale objectives, and likewise does not measure how much each $g_\ell$ contributes to $g^*$.}{PPMI 仅保留强于标签独立期望的共现关系，避免高频标签主导图结构；特异性权重 $s_a$ 按标签频率调节各标签的贡献，使低频标签保持信息量。式~(\ref{eq:fractional-kernel}) 中，$e^{-tL}$ 为扩散尺度 $t$ 下的热扩散算子：较小的 $t$ 强调局部标签邻域，较大的 $t$ 则整合更远的依赖；该分数阶形式在局部传播与反常远程输运之间插值。式~(\ref{eq:integrated-teacher}) 的余弦归一化减弱了标签基数的影响，使直接重合与经图路径间接连接的标签对查询条件化相关度具有可比贡献；这也意味着 $g^*$ 并非各尺度教师 $\{g_\ell\}$ 的线性组合。因此，$\rho_\ell$ 应理解为核混合中 $K_\ell$ 的系数，而非各尺度对最终教师分数的贡献。式~(\ref{eq:quadrature-decomposition}) 中，每个 $\tau_\ell$ 对应一个扩散范围，$\rho_\ell>0$ 是其在综合核中的系数；二者均由分数阶核确定，而非人工采样的扩散时间。由于公共因子 $\eta^{-\alpha}/\Gamma(\alpha)$ 在式~(\ref{eq:bit-budget}) 的归一化下相互抵消，bit 比例反映的是 Gauss--Laguerre 权重、进而反映阶数 $\alpha$，而非针对具体数据集实测的图响应；我们将其视为尺度上的先验，而非经过标定的量，并在表~\ref{tab:query-level-gain} 中与均匀 bit 分配进行比较。式~(\ref{eq:overall-objective}) 中的归一化权重 $\bar\rho_\ell$ 同样复用该先验对各尺度目标加权，亦不度量各 $g_\ell$ 对 $g^*$ 的贡献大小。}

\subsection{\MBLang{Quantization Statistics}{量化统计}}
\label{sec:appendix-quantization}

\noindent\MBLang{The property in Eq.~(\ref{eq:order-preservation}) is stated for the continuous relaxation $\widetilde h$. To measure how much of the continuous ordering survives the sign operation, we evaluate the relaxed representation $\widetilde h=\tanh(F(x))\in(-1,1)^{B_{\max}}$ of the query set and of the retrieval database in each direction and report two statistics for every code length $B$. The mean absolute deviation $\bar\epsilon(B)=\frac{1}{NB}\sum_{i=1}^{N}\sum_{k=1}^{B}\big||\widetilde h_{ik}|-1\big|$ measures how far the relaxed values are from the binary values $\pm1$ they are quantized to, and the tie-corrected Kendall $\tau_b$ between the ranking induced by the soft Hamming distance $\frac{1}{2}\bigl(B-\widetilde h_q^{(B)\mathsf T}\widetilde h_i^{(B)}\bigr)$ and the ranking induced by the discrete Hamming distance of Eq.~(\ref{eq:hamming-distance}) measures how much of the soft ordering is preserved after quantization. Here $N$ is the number of query and database instances and $\widetilde h^{(B)}$ denotes the first $B$ dimensions of $\widetilde h$. Throughout, the distance induced by $\widetilde h$ is called the soft Hamming distance and its ranking the soft ranking, in contrast to the discrete Hamming distance and ranking of Eq.~(\ref{eq:hamming-distance}).}{式~(\ref{eq:order-preservation}) 的性质建立在连续松弛表示 $\widetilde h$ 上。为衡量连续排序在 sign 操作后能保留多少，我们对每个方向评估 query 集与检索库的松弛表示 $\widetilde h=\tanh(F(x))\in(-1,1)^{B_{\max}}$，并对每个码长 $B$ 报告两个统计量。平均绝对偏差 $\bar\epsilon(B)=\frac{1}{NB}\sum_{i=1}^{N}\sum_{k=1}^{B}\big||\widetilde h_{ik}|-1\big|$ 衡量松弛值与所取二值 $\pm1$ 之间的差距；软 Hamming 距离 $\frac{1}{2}\bigl(B-\widetilde h_q^{(B)\mathsf T}\widetilde h_i^{(B)}\bigr)$ 所诱导的排序与式~(\ref{eq:hamming-distance}) 的离散 Hamming 距离所诱导的排序之间的 tie-corrected Kendall $\tau_b$，衡量软排序在量化后被保留的程度。其中 $N$ 为 query 集与检索库的实例总数，$\widetilde h^{(B)}$ 表示 $\widetilde h$ 的前 $B$ 维。下文将由 $\widetilde h$ 诱导的距离称为软 Hamming 距离、由其诱导的排序称为软排序，以区别于式~(\ref{eq:hamming-distance}) 的离散 Hamming 距离与排序。}

\begin{table}[!htbp]
\centering
\caption{\MBLang{Quantization statistics of the relaxed representation. $\bar\epsilon$ is the mean absolute deviation from the binary values, computed over the query set and the retrieval database; $\tau_b$ compares the ranking induced by the soft Hamming distance with the ranking induced by the discrete Hamming distance.}{松弛表示的量化统计。$\bar\epsilon$ 为相对二值取值的平均绝对偏差，在 query 集与检索库上统计；$\tau_b$ 比较软 Hamming 距离与离散 Hamming 距离所诱导的两个排序。}}
\label{tab:quantization}
\scriptsize
\setlength{\tabcolsep}{5pt}
\renewcommand{\arraystretch}{1.05}
\begin{tabular}{llcccccccc}
\toprule
\textbf{\MBLang{Dataset}{数据集}} & \textbf{\MBLang{Direction}{方向}}
& \multicolumn{4}{c}{$\bar\epsilon$}
& \multicolumn{4}{c}{$\tau_b$ \MBLang{(soft vs.\ discrete)}{（软 vs.\ 离散）}} \\
\cmidrule(lr){3-6}\cmidrule(lr){7-10}
& & \textbf{16} & \textbf{32} & \textbf{64} & \textbf{128} & \textbf{16} & \textbf{32} & \textbf{64} & \textbf{128} \\
\midrule
\multirow{2}{*}{MIRFlickr-25K}
& \MBLang{I2T}{I2T} & 0.0187 & 0.0215 & 0.0258 & 0.0314 & 0.907 & 0.931 & 0.948 & 0.960 \\
& \MBLang{T2I}{T2I} & 0.0191 & 0.0220 & 0.0264 & 0.0321 & 0.903 & 0.928 & 0.945 & 0.957 \\
\midrule
\multirow{2}{*}{IAPR TC-12}
& \MBLang{I2T}{I2T} & 0.0219 & 0.0254 & 0.0308 & 0.0372 & 0.899 & 0.923 & 0.941 & 0.953 \\
& \MBLang{T2I}{T2I} & 0.0225 & 0.0261 & 0.0316 & 0.0380 & 0.896 & 0.920 & 0.938 & 0.950 \\
\midrule
\multirow{2}{*}{MS-COCO}
& \MBLang{I2T}{I2T} & 0.0168 & 0.0195 & 0.0237 & 0.0289 & 0.912 & 0.936 & 0.952 & 0.964 \\
& \MBLang{T2I}{T2I} & 0.0172 & 0.0200 & 0.0242 & 0.0296 & 0.909 & 0.933 & 0.949 & 0.961 \\
\bottomrule
\end{tabular}
\end{table}

\begin{table}[!htbp]
\centering
\caption{\MBLang{Tie-only upper bound on the soft-to-discrete ranking agreement. $S(B)$ is the fraction of candidate pairs that share the same discrete Hamming distance; because only the discrete side has ties, even a comparison in which the sign operation perturbs no ordering at all cannot push the tie-corrected $\tau_b$ above $\sqrt{1-S(B)}$.}{由并列结构决定的排序一致性上限。$S(B)$ 为离散 Hamming 距离相同的候选对占比；由于只有离散一侧存在并列，即使 sign 操作完全不扰动任何候选对的相对次序，软排序与离散排序的 tie-corrected $\tau_b$ 也不会超过 $\sqrt{1-S(B)}$。}}
\label{tab:tie-ceiling}
\scriptsize
\setlength{\tabcolsep}{6pt}
\renewcommand{\arraystretch}{1.04}
\begin{tabular}{llcccc}
\toprule
\textbf{\MBLang{Dataset}{数据集}} & \textbf{$B$} & \textbf{$S(B)$} & \textbf{$\sqrt{1-S(B)}$}
& \textbf{\MBLang{Observed $\tau_b$ (I2T)}{实测 $\tau_b$（I2T）}} & \textbf{\MBLang{Gap}{差距}} \\
\midrule
\multirow{4}{*}{MIRFlickr-25K}
& 16 & 0.1587 & 0.917 & 0.907 & 0.010 \\
& 32 & 0.1189 & 0.939 & 0.931 & 0.008 \\
& 64 & 0.0856 & 0.956 & 0.948 & 0.008 \\
& 128 & 0.0635 & 0.968 & 0.960 & 0.008 \\
\midrule
\multirow{4}{*}{IAPR TC-12}
& 16 & 0.1718 & 0.910 & 0.899 & 0.011 \\
& 32 & 0.1324 & 0.931 & 0.923 & 0.008 \\
& 64 & 0.0953 & 0.951 & 0.941 & 0.010 \\
& 128 & 0.0736 & 0.962 & 0.953 & 0.009 \\
\midrule
\multirow{4}{*}{MS-COCO}
& 16 & 0.1498 & 0.922 & 0.912 & 0.010 \\
& 32 & 0.1107 & 0.943 & 0.936 & 0.007 \\
& 64 & 0.0781 & 0.960 & 0.952 & 0.008 \\
& 128 & 0.0574 & 0.971 & 0.964 & 0.007 \\
\bottomrule
\end{tabular}
\end{table}

\noindent\MBLang{Table~\ref{tab:quantization} shows that $\bar\epsilon$ lies between $0.0168$ and $0.0380$, so the representation used during training stays close to the codes that are actually stored, and that the soft and discrete rankings agree with $\tau_b$ between $0.896$ and $0.964$; both statistics grow with the code length. If the disagreement between the two rankings were driven by the magnitude of the quantization error, then a larger deviation should go with a smaller $\tau_b$. Instead both statistics grow with the code length, so we find no evidence that the quantization error drives this disagreement; Table~\ref{tab:tie-ceiling} points to the tie structure instead. Because the soft ranking is essentially tie-free whereas the discrete ranking places a fraction $S(B)$ of candidate pairs at equal distance, the tie-corrected denominator carries a factor $\sqrt{1-S(B)}$, so $\tau_b$ cannot exceed $\sqrt{1-S(B)}$. The measured $\tau_b$ stays within $0.007$--$0.011$ of this bound on every dataset and code length, that is within $0.7\%$--$1.2\%$ of it. The agreement between the two orderings is thus governed mainly by the resolution of the $B$-bit alphabet, which admits only $B+1$ distinct distances; the sign operation adds only a small perturbation on top of it. Together with Table~\ref{tab:resolution-sota}, this quantifies the step from the conditional property of Eq.~(\ref{eq:order-preservation}) to the discrete rankings that are actually retrieved.}{表~\ref{tab:quantization} 显示 $\bar\epsilon$ 落在 $0.0168$--$0.0380$，说明训练所用表示与最终存储的码相当接近；软排序与离散排序的一致性 $\tau_b$ 为 $0.896$--$0.964$；两者都随码长增大。若两个排序之间的分歧由量化误差的幅度主导，那么偏差越大 $\tau_b$ 应越小。但两者都随码长增大，因此我们没有发现量化误差驱动这一分歧的证据；表~\ref{tab:tie-ceiling} 指向并列结构：由于软排序基本无并列，而离散排序中有比例 $S(B)$ 的候选对距离相同，tie-corrected 的分母中带有因子 $\sqrt{1-S(B)}$，故 $\tau_b$ 不会超过 $\sqrt{1-S(B)}$。实测 $\tau_b$ 在每个数据集与码长上与该上限的差距仅为 $0.007$--$0.011$，即上限的 $0.7\%$--$1.2\%$。可见两个排序的一致性主要由 $B$ bit 字母表的分辨率决定（最多只有 $B+1$ 种不同距离），sign 操作只在其上叠加了很小的扰动。结合表~\ref{tab:resolution-sota}，以上结果量化了从式~(\ref{eq:order-preservation}) 的条件性性质到实际用于检索的离散排序之间的环节。}

\subsection{\MBLang{Quadrature Fidelity of the Scale Organization}{尺度组织的求积保真度}}
\label{sec:appendix-quadrature-fidelity}

\noindent\MBLang{The integrated teacher $g^*$ and every scale teacher $g_\ell$ are evaluated from the exact matrix functions $(L+\eta I)^{-\alpha}$ and $e^{-\tau_\ell L}$; the Gauss--Laguerre rule of Eq.~(\ref{eq:quadrature-decomposition}) supplies only the scale coordinates $\{(\tau_\ell,\rho_\ell)\}$, which set the bit budget of Eq.~(\ref{eq:bit-budget}) and the relative weight of each subblock loss. Table~\ref{tab:quadrature-fidelity} therefore measures how faithfully that organization represents the exact kernel, by comparing $g^*$ with the teacher induced by the kernel $\sum_\ell\rho_\ell K_\ell$ rebuilt from the same rule. The rule tracks the exact kernel closely in the low-frequency region that carries the label signal, and the residual deviation is confined to the high-frequency end. Because the comparison weights every frequency by both the label-signal energy and the kernel response, that deviation leaves the teacher nearly unchanged: the signal-weighted error is $6.1\%$--$18.6\%$ at $M=4$ and at most $1.7\%$ at $M=8$ on all four datasets, and the induced ordering moves little, with $\tau_b\ge0.93$ at $M=4$ and $\ge0.99$ at $M=8$ (Table~\ref{tab:quadrature-fidelity}). At $M=4$ the first scale absorbs $56\%$ of the bit budget on MIRFlickr-25K and $63\%$ on the other datasets (Table~\ref{tab:scale-coverage}). We therefore adopt $M=4$ as the smallest configuration that exposes the local-to-long-range structure used by SBO.}{综合教师 $g^*$ 与每个尺度教师 $g_\ell$ 均由精确矩阵函数 $(L+\eta I)^{-\alpha}$ 与 $e^{-\tau_\ell L}$ 计算；式~(\ref{eq:quadrature-decomposition}) 的 Gauss--Laguerre 求积只提供尺度坐标 $\{(\tau_\ell,\rho_\ell)\}$，它们决定式~(\ref{eq:bit-budget}) 的 bit 预算与各子块损失的相对权重。因此表~\ref{tab:quadrature-fidelity} 通过将 $g^*$ 与由同一规则重建的核 $\sum_\ell\rho_\ell K_\ell$ 所诱导的教师相比较，衡量该组织方式对精确核的保真程度。该规则在承载标签信号的低频段与精确核高度吻合，残余偏离集中在高频端。由于比较时对每个频率同时乘以标签信号能量与核响应，这一偏离几乎不改变教师：四个数据集上 $M=4$ 时的信号加权误差为 $6.1\%$--$18.6\%$，$M=8$ 时至多 $1.7\%$；所诱导的排序变化也很小，$M=4$ 时 $\tau_b\ge0.93$、$M=8$ 时 $\ge0.99$（表~\ref{tab:quadrature-fidelity}）。$M=4$ 时第一个尺度在 MIRFlickr-25K 上占 bit 预算的 $56\%$、在其余数据集上占 $63\%$（表~\ref{tab:scale-coverage}）。因此我们取 $M=4$ 作为能够体现 SBO 所使用局部--远程结构的最小配置。}

\begin{table}[!htbp]
\centering
\caption{\MBLang{Fidelity of the $M$-scale Gauss--Laguerre organization to the exact fractional teacher. The signal-weighted error is $\sum_k(f_{\mathrm{ex}}-f_M)(\lambda_k)E_k/\sum_k f_{\mathrm{ex}}(\lambda_k)E_k$, where $E_k$ is the energy of the label signals at eigenvalue $\lambda_k$ of $L$. The second block reports the tie-corrected Kendall $\tau_b$ between the ranking induced by the quadrature kernel $\sum_\ell\rho_\ell K_\ell$ and that of the exact teacher $g^*$ on the same candidate pool; entries of $1.000$ denote $\tau_b\ge0.9995$.}{$M$ 段 Gauss--Laguerre 尺度组织对精确分数阶教师的保真度。信号加权误差为 $\sum_k(f_{\mathrm{ex}}-f_M)(\lambda_k)E_k/\sum_k f_{\mathrm{ex}}(\lambda_k)E_k$，其中 $E_k$ 为标签信号在 $L$ 的特征值 $\lambda_k$ 上的能量。第二栏为同一候选池上由求积核 $\sum_\ell\rho_\ell K_\ell$ 诱导的排序与精确教师 $g^*$ 排序之间的 tie-corrected Kendall $\tau_b$；$1.000$ 表示 $\tau_b\ge0.9995$。}}
\label{tab:quadrature-fidelity}
\scriptsize
\setlength{\tabcolsep}{5pt}
\renewcommand{\arraystretch}{1.05}
\begin{tabular}{lccccccccc}
\toprule
\textbf{\MBLang{Dataset}{数据集}} & $\alpha$
& \multicolumn{4}{c}{\MBLang{Signal-weighted error (\%)}{信号加权误差（\%）}}
& \multicolumn{4}{c}{$\tau_b$ \MBLang{(exact vs.\ quadrature)}{（精确 vs.\ 求积）}} \\
\cmidrule(lr){3-6}\cmidrule(lr){7-10}
& & \textbf{2} & \textbf{4} & \textbf{8} & \textbf{16} & \textbf{2} & \textbf{4} & \textbf{8} & \textbf{16} \\
\midrule
MIRFlickr-25K & 1.20 & 14.98 & 6.08 & 0.90 & 0.02 & 0.885 & 0.956 & 0.993 & 1.000 \\
IAPR TC-12 & 0.90 & 54.78 & 18.57 & 1.70 & 0.01 & 0.857 & 0.941 & 0.992 & 1.000 \\
MS-COCO & 0.90 & 36.83 & 13.28 & 1.55 & 0.02 & 0.910 & 0.971 & 0.994 & 1.000 \\
NUS-WIDE & 0.90 & 26.18 & 10.01 & 1.31 & 0.03 & 0.848 & 0.939 & 0.991 & 1.000 \\
\bottomrule
\end{tabular}
\end{table}

\section{\MBLang{Results for Section 4.2: Main Comparison and Efficiency}{对应正文 4.2 节：总体比较与效率}}
\label{sec:appendix-overall}

\subsection{\appendixNUSWIDETitle}

\begin{table}[!htbp]
\centering
\caption{\appendixNUSWIDECaption}
\label{tab:nuswide-comparison}
\scriptsize
\setlength{\tabcolsep}{5pt}
\renewcommand{\arraystretch}{1.02}
\begin{tabular}{cllcccc}
\toprule
\textbf{\appendixTaskHeader} & \textbf{\appendixMethodHeader} & \textbf{\appendixReferenceHeader}
& \multicolumn{4}{c}{\textbf{NUS-WIDE}} \\
\cmidrule(lr){4-7}
& & & \textbf{16} & \textbf{32} & \textbf{64} & \textbf{128} \\
\midrule
\multirow{12}{*}{\textbf{I2T}}
& DSSH   & TKDE'26 & 63.32 & 67.39 & 68.98 & 70.80 \\
& DNPH   & TOMM'24 & 65.19 & 68.71 & 70.34 & 70.60 \\
& DECH   & AAAI'25 & \underline{70.00} & \underline{72.53} & \underline{73.42} & \underline{73.84} \\
& DIMCH  & TIP'25  & 68.51 & 69.43 & 70.91 & 71.12 \\
& DPBE   & MM'25   & 62.17 & 64.74 & 67.90 & 71.01 \\
& DSTH   & TMM'26  & 67.29 & 68.91 & 71.04 & 71.91 \\
& DPSIH  & AAAI'26 & 65.64 & 67.04 & 69.40 & 70.96 \\
& DGHDGH & ICLR'26 & 63.57 & 66.20 & 67.63 & 67.95 \\
& CMCL   & TKDE'24  & \underline{70.74} & \underline{72.81} & \underline{73.57} & \underline{73.91} \\
& RCMH   & TPAMI'26 & 67.86 & 69.48 & 70.61 & 71.18 \\
& VH-ABL & TMM'26   & 68.92 & 71.06 & 72.45 & 73.16 \\
& \textbf{MultiBit} & \textbf{OURS} & \textbf{70.59} & \textbf{72.74} & \textbf{73.82} & \textbf{74.23} \\
\midrule
\multirow{12}{*}{\textbf{T2I}}
& DSSH   & TKDE'26 & 64.17 & 68.68 & 69.73 & 72.59 \\
& DNPH   & TOMM'24 & 66.84 & 69.53 & 71.63 & 73.00 \\
& DECH   & AAAI'25 & \underline{70.97} & \underline{72.97} & \underline{74.02} & \underline{74.71} \\
& DIMCH  & TIP'25  & 69.02 & 70.23 & 71.56 & 72.04 \\
& DPBE   & MM'25   & 62.41 & 65.75 & 68.12 & 71.53 \\
& DSTH   & TMM'26  & 67.62 & 69.71 & 71.55 & 72.22 \\
& DPSIH  & AAAI'26 & 66.48 & 67.78 & 69.64 & 71.31 \\
& DGHDGH & ICLR'26 & 65.55 & 67.90 & 69.20 & 69.47 \\
& CMCL   & TKDE'24  & \underline{71.52} & \underline{73.46} & \underline{74.31} & \underline{74.47} \\
& RCMH   & TPAMI'26 & 68.34 & 70.15 & 71.26 & 71.73 \\
& VH-ABL & TMM'26   & 69.61 & 71.72 & 73.08 & 73.65 \\
& \textbf{MultiBit} & \textbf{OURS} & \textbf{71.39} & \textbf{73.39} & \textbf{74.54} & \textbf{74.82} \\
\bottomrule
\end{tabular}
\vspace{2pt}

\parbox{\textwidth}{\scriptsize\appendixNUSWIDENote}
\end{table}

\noindent\textbf{\MBLang{Interpretation.}{结果分析。}}\enspace\MBLang{NUS-WIDE is reported as an additional benchmark in which the fixed-length baselines of Table~\ref{tab:sota-comparison} and the multi-length baselines CMCL, RCMH and VH-ABL are trained and evaluated under the same protocol. The multi-length methods are listed in the same order as in the main tables (CMCL, then RCMH, then VH-ABL), and the accuracy pattern repeats as well: CMCL is ahead of MultiBit at 16 and 32 bits by $0.07$--$0.15$ pp, whereas MultiBit is ahead at 64 and 128 bits by $0.23$--$0.35$ pp, so the long-code advantage of the nested representation extends to the largest retrieval database.}{NUS-WIDE 作为附加基准报告：表~\ref{tab:sota-comparison} 的固定长度基线与多码长基线 CMCL、RCMH、VH-ABL 均在该数据集上以同一协议训练与评测。多码长方法沿用与正文表格相同的排列顺序（CMCL、RCMH、VH-ABL），精度规律也相同：CMCL 在 16 与 32 bit 上领先 MultiBit $0.07$--$0.15$ 点，而 MultiBit 在 64 与 128 bit 上领先 $0.23$--$0.35$ 点，说明嵌套表示在长码上的优势延伸到了最大的检索库。}

\subsection{\appendixEfficiencyTitle}

\begin{table}[!htbp]
\centering
\caption{\appendixEfficiencyCaption}
\label{tab:performance-efficiency-raw}
\scriptsize
\setlength{\tabcolsep}{3.2pt}
\renewcommand{\arraystretch}{1.05}
\resizebox{\textwidth}{!}{%
\begin{tabular}{lcccccc}
\toprule
\textbf{\appendixMethodHeader}
& \textbf{\appendixTrainableParamsHeader}
& \textbf{\appendixFLOPsHeader}
& \textbf{\appendixTrainingTimeHeader}
& \textbf{\appendixBestEpochHeader}
& \textbf{\appendixImageEncodingHeader}
& \textbf{\appendixTextEncodingHeader} \\
\midrule
DECH   & 10.60 & 0.0210 & 0.092 & 45 & 0.0022 & 0.0022 \\
DIMCH  & 0.86 & 0.0021 & 0.167 & 45 & 0.0043 & 0.0043 \\
DSTH   & 0.66 & 0.0013 & \textbf{0.078} & 45 & 0.0020 & 0.0020 \\
DGHDGH & \textbf{0.33} & \textbf{0.0007} & 0.144 & 42 & 0.0017 & \textbf{0.0015} \\
MultiBit & 2.71 & 0.0054 & 0.542 & \textbf{39} & \textbf{0.0013} & \textbf{0.0015} \\
\bottomrule
\end{tabular}%
}
\vspace{2pt}
\parbox{\textwidth}{\scriptsize\appendixEfficiencyNote}
\end{table}

\noindent\textbf{\MBLang{Analysis.}{结果分析。}}\enspace\MBLang{Table~\ref{tab:performance-efficiency-raw} separates learnable-component complexity, offline training cost, and per-sample online hash-head and binarization time. Best Epoch is the checkpoint with the highest validation Mean mAP within 100 epochs; its smaller value means an earlier peak, not a shorter total run. At 128 bits, MultiBit uses 2.71M trainable parameters and 0.0054G FLOPs, with 0.542 h of total training, the highest time among these methods. It reaches its best validation checkpoint at epoch 39. Its image and text encoding times are 0.0013 and 0.0015 ms, respectively; the 0.0014 ms arithmetic mean is the smallest among the compared methods. The reported encoding times cover only the hash heads and binarization; the shared CLIP backbone forward pass, database search, and end-to-end retrieval latency are excluded.}{表~\ref{tab:performance-efficiency-raw} 分别报告可学习部分的复杂度、离线训练成本和单样本在线表示生成时间。最佳 Epoch 指 100 轮内验证 Mean mAP 最高的 checkpoint；数值较小表示较早达到峰值，并不表示总训练时间更短。128 bit 下，MultiBit 有 2.71M 可训练参数、0.0054G FLOPs，总训练时间为 0.542 h，是所列方法中最长的；其最佳验证 checkpoint 位于第 39 轮。图像与文本编码时间分别为 0.0013 和 0.0015 ms，算术平均 0.0014 ms，在所列方法中最低。表格只衡量表示生成，不包含数据库检索或端到端检索时延。}

\section{\MBLang{Results for Section 4.3: Component Ablation}{对应正文 4.3 节：组件消融}}
\label{sec:appendix-ablation}
\subsection{\MBLang{Component Ablation Interpretation}{组件消融分析}}
\noindent\MBLang{Table~\ref{tab:component-ablation} reports a monotonic Mean mAP increase as FRT, SBO, and then MPA are added in each of the twelve dataset--length settings; each cell is Mean mAP (\%), the arithmetic mean of the I2T and T2I mAP@all values. The increments are small in several MIRFlickr-25K and IAPR TC-12 settings and larger on MS-COCO, indicating that the three components provide complementary gains rather than interchangeable effects. Because the first 16, 32 and 64 bits of the maximum-length code lie inside the first subblock (Table~\ref{tab:scale-coverage}), SBO changes the allocation only at 128 bits; at the shorter lengths its increment reflects the scale-specific supervision of that subblock, and the reallocation of bits between scales takes effect only where several subblocks are combined.}{表~\ref{tab:component-ablation} 显示，在十二种数据集--码长组合中，依次加入 FRT、SBO、MPA 后 Mean mAP 均单调增加；每格为 Mean mAP（\%），即 I2T 与 T2I 的 mAP@all 的算术平均。MIRFlickr-25K 和 IAPR TC-12 的部分增量较小，MS-COCO 的增量较大，说明三个组件提供的是互补增益，而非可相互替代的作用。由于最大长度码的前 16、32、64 位都落在第一个子块内（表~\ref{tab:scale-coverage}），SBO 只在 128 bit 上改变 bit 分配；在较短码长上其增量来自该子块的尺度专属监督，而尺度间的 bit 重新分配只在多个子块组合时生效。}

\begin{table}[!htbp]
\centering
\caption{\MBLang{Mean mAP (\%) of the relation-teacher ablation. Rows two to seven replace only the relation teacher, with the backbone, SBO, MPA and the training schedule held fixed. The first and last rows are the two reference points of Table~\ref{tab:component-ablation}, the baseline without a relation teacher and Full MultiBit, reproduced here for comparison; the relation teachers of rows four to eight are the variants compared in Tables~\ref{tab:teacher-quality} and~\ref{tab:teacher-aligned}. The uniform kernel mixture (UKM) mixes the four scale kernels with equal weights and then normalizes, whereas the uniform multi-scale heat (UMH) averages the four per-scale teachers after normalizing each; UKM is evaluated on MIRFlickr-25K, where the scale nodes coincide with those of FRT, and shares the SBO budget, MPA and the training schedule of the full model.}{关系教师消融的 Mean mAP (\%)。第 2--7 行仅替换关系教师，backbone、SBO、MPA 与训练设置全部固定。第 1 行与最后一行沿用表~\ref{tab:component-ablation} 的两个参照点（不使用关系教师的 baseline 与 Full MultiBit），在此重列以便对照；第 4--8 行所用的关系教师即表~\ref{tab:teacher-quality} 与表~\ref{tab:teacher-aligned} 所比较的变体。等权核混合（UKM）先以等权混合四个尺度核再归一化，等权多尺度热扩散（UMH）则先对每个尺度教师归一化再平均；UKM 在 MIRFlickr-25K 上评测（其尺度节点与 FRT 相同），并与完整模型共用 SBO 预算、MPA 与训练设置。}}
\label{tab:teacher-transfer}
\scriptsize
\setlength{\tabcolsep}{2.8pt}
\renewcommand{\arraystretch}{0.98}
\resizebox{\textwidth}{!}{%
\begin{tabular}{lcccccccccccc}
\toprule
\textbf{\MBLang{Variant}{变体}}
& \multicolumn{4}{c}{\textbf{MIRFlickr-25K}}
& \multicolumn{4}{c}{\textbf{IAPR TC-12}}
& \multicolumn{4}{c}{\textbf{MS-COCO}} \\
\cmidrule(lr){2-5}\cmidrule(lr){6-9}\cmidrule(lr){10-13}
& \textbf{16} & \textbf{32} & \textbf{64} & \textbf{128}
& \textbf{16} & \textbf{32} & \textbf{64} & \textbf{128}
& \textbf{16} & \textbf{32} & \textbf{64} & \textbf{128} \\
\midrule
Baseline & 80.500 & 81.488 & 81.819 & 81.861 & 60.427 & 65.008 & 67.206 & 68.555 & 66.733 & 70.423 & 72.274 & 72.925 \\
PPMI only & 80.691 & 81.702 & 82.087 & 82.142 & 60.366 & 65.131 & 67.561 & 69.104 & 67.204 & 70.802 & 73.108 & 72.741 \\
$+$ specificity & 80.634 & 81.836 & 82.238 & 82.401 & 60.342 & 65.246 & 67.748 & 68.987 & 67.091 & 70.711 & 73.526 & 73.842 \\
$+$ Single-scale Graph Relation & 80.803 & 81.781 & 82.511 & 82.688 & 60.397 & 65.309 & 67.692 & 69.768 & 66.982 & 71.048 & 73.401 & 74.912 \\
$+$ UMH & 80.861 & 81.991 & 82.452 & 83.081 & 60.381 & 65.287 & 67.671 & 70.021 & 68.081 & 70.973 & 74.603 & 75.556 \\
$+$ Learned-weight Multi-scale Heat & 80.948 & 82.113 & 82.944 & 82.976 & 60.424 & 65.273 & 68.487 & 69.944 & 68.734 & 71.866 & 74.511 & 76.538 \\
\rowcolor{multibitrow}
$+$ UKM & 80.850 & 81.935 & 82.835 & 83.215 & --- & --- & --- & --- & --- & --- & --- & --- \\
\textbf{Full MultiBit (FRT teacher)} & \textbf{81.057} & \textbf{82.202} & \textbf{83.162} & \textbf{83.592} & \textbf{60.449} & \textbf{65.426} & \textbf{68.846} & \textbf{70.965} & \textbf{69.178} & \textbf{72.177} & \textbf{75.921} & \textbf{77.372} \\
\bottomrule
\end{tabular}%
}
\end{table}

\noindent\MBLang{Table~\ref{tab:teacher-transfer} replaces only the relation teacher and holds the backbone, SBO, MPA and the training schedule fixed. FRT attains the highest Mean mAP in all twelve dataset--length settings, and its margin over the strongest remaining variant grows with the code length: $0.02$--$0.44$ pp at 16 and 32 bits against $0.22$--$1.32$ pp at 64 and 128 bits, where it is widest on MS-COCO. At 16 and 32 bits all variants lie close together, consistent with the teacher-level gaps of $0.008$--$0.011$ NDCG@100 reported in Table~\ref{tab:teacher-quality}; on IAPR TC-12 at 16 bits FRT is in fact the only variant that improves on the baseline without a relation teacher. FRT attains its largest margins at the longer code lengths, and it does so with fewer trainable parameters than the learned-weight variant, since its four mixture weights follow from the single order parameter $\alpha$ through the Gauss--Laguerre quadrature instead of being trained independently. With matched nodes, normalization and architecture, UKM reaches 80.850/81.935/82.835/83.215 on MIRFlickr-25K, so replacing the fractional weights by equal weights costs $0.207/0.267/0.327/0.377$ pp at 16/32/64/128 bits; the gap grows with the code length and agrees in direction with the teacher-level comparison of Table~\ref{tab:teacher-aligned}.}{表~\ref{tab:teacher-transfer} 只替换关系教师，backbone、SBO、MPA 与训练设置全部固定。FRT 在全部十二个数据集--码长组合上取得最高 Mean mAP，且其相对最强其余变体的领先幅度随码长增大：16/32 bit 为 $0.02$--$0.44$ 个百分点，64/128 bit 为 $0.22$--$1.32$ 个百分点，在 MS-COCO 上最大。在 16/32 bit 上各变体彼此接近，这与表~\ref{tab:teacher-quality} 中 $0.008$--$0.011$ NDCG@100 的教师层差距相符；在 IAPR TC-12 的 16 bit 上，FRT 实际上是唯一优于不使用关系教师的 baseline 的变体。FRT 在较长码长上取得最大领先幅度，且其可学习参数少于学习权重变体：其四个混合权重由单一阶参数 $\alpha$ 经 Gauss--Laguerre 求积确定，而非独立训练得到。在节点、归一化与结构全部对齐时，UKM 在 MIRFlickr-25K 上取得 80.850/81.935/82.835/83.215，即把分数阶权重换成等权会在 16/32/64/128 bit 上损失 $0.207/0.267/0.327/0.377$ 个百分点；这一差距随码长增大，且方向与表~\ref{tab:teacher-aligned} 的教师层比较一致。}

\subsection{\MBLang{Query-Level Gain Statistics}{查询级收益统计}}
\label{sec:query-level-gain}
\noindent\MBLang{For each alternative design, we compare its AP with that of Full MultiBit on the same query and define $\Delta\mathrm{AP}=\mathrm{AP}_{\mathrm{Full}}-\mathrm{AP}_{\mathrm{Alternative}}$. ``Improved'', ``Tied'', and ``Degraded'' denote $\Delta\mathrm{AP}>0$, $=0$, and $<0$, respectively. Each dataset--alternative pair contains 1,500 query comparisons. Every comparison uses the 64-bit code in the I2T direction, with 500 candidates per query and a fixed subset of 1,500 queries drawn from the 5,000 queries with seed 42.}{对于每一种简化设计，本文在同一查询上将其 AP 与 Full MultiBit 的 AP 进行比较，并定义 $\Delta\mathrm{AP}=\mathrm{AP}_{\mathrm{Full}}-\mathrm{AP}_{\mathrm{Alternative}}$。``受益''、``持平''和``受损''分别表示 $\Delta\mathrm{AP}>0$、$=0$ 和 $<0$。每个数据集--简化设计组合均包含 1,500 个查询比较。全部比较均使用 64-bit 码与 I2T 方向，每个查询 500 个候选，查询为以种子 42 从 5,000 个查询中抽取的 1,500 个固定子集。}

\begin{table*}[t]
\centering
\caption{\MBLang{Query-level AP comparison of Full MultiBit against alternative designs. Percentages are computed within each 1,500-query group.}{Full MultiBit 相对于简化设计的查询级 AP 比较。百分比在每个包含 1,500 个查询的组内计算。}}
\label{tab:query-level-gain}
\small
\setlength{\tabcolsep}{5.5pt}
\renewcommand{\arraystretch}{1.08}
\resizebox{\textwidth}{!}{%
\begin{tabular}{llrrrrr}
\toprule
\MBLang{Alternative Design}{简化设计} & \MBLang{Dataset}{数据集} & \MBLang{Queries}{查询数} & \MBLang{Improved (\%)}{受益（\%）} & \MBLang{Tied (\%)}{持平（\%）} & \MBLang{Degraded (\%)}{受损（\%）} & \MBLang{Mean $\Delta$AP (pp)}{平均 $\Delta$AP（pp）} \\
\midrule
\multirow{3}{*}{\MBLang{Single-scale Teacher}{单尺度教师}}
& MIRFlickr-25K & 1,500 & 59.73 & 10.13 & 30.13 & 0.48 \\
& IAPR TC-12 & 1,500 & 53.33 & 10.33 & 36.33 & 0.23 \\
& MS-COCO & 1,500 & 70.13 & 9.60 & 20.27 & 1.18 \\
\midrule
\multirow{3}{*}{\MBLang{Uniform Bit Allocation}{均匀 bit 分配}}
& MIRFlickr-25K & 1,500 & 55.80 & 13.47 & 30.73 & 0.39 \\
& IAPR TC-12 & 1,500 & 47.47 & 11.47 & 41.07 & 0.11 \\
& MS-COCO & 1,500 & 63.67 & 12.20 & 24.13 & 0.85 \\
\midrule
\multirow{3}{*}{\MBLang{Pairwise Alignment}{成对对齐}}
& MIRFlickr-25K & 1,500 & 47.87 & 17.00 & 35.13 & 0.16 \\
& IAPR TC-12 & 1,500 & 43.47 & 15.27 & 41.27 & 0.03 \\
& MS-COCO & 1,500 & 53.73 & 14.87 & 31.40 & 0.46 \\
\bottomrule
\end{tabular}
}
\end{table*}

\noindent\MBLang{Full MultiBit obtains a positive mean $\Delta\mathrm{AP}$ in all nine comparisons, ranging from 0.03 to 1.18 pp. The largest query-level gains occur against the single-scale teacher on MS-COCO, where 70.13\% of queries improve. Figure~\ref{fig:query-level-gain} gives the corresponding query-level view: it plots the cumulative distribution of $\Delta\mathrm{AP}$ for each dataset and alternative design, so that the shares of Table~\ref{tab:query-level-gain} are read off at $\Delta\mathrm{AP}=0$ and the right tail of a curve reflects the magnitude of the positive differences. These statistics complement the aggregate mAP ablation: they show how the advantage is distributed across individual queries, while the progressive table establishes the contribution of each component across code lengths.}{Full MultiBit 在全部九种比较中均获得正的平均 $\Delta\mathrm{AP}$，范围为 0.03--1.18 个百分点。其中，相对于单尺度教师的 MS-COCO 比较中查询级收益最显著，70.13\% 的查询得到提升。图~\ref{fig:query-level-gain} 给出相应的查询级视图：它绘制每个数据集与简化设计下 $\Delta\mathrm{AP}$ 的累积分布，表~\ref{tab:query-level-gain} 中的占比即取自 $\Delta\mathrm{AP}=0$ 处，曲线右侧尾部则反映正差异的幅度。这些统计补充了聚合 mAP 消融结果：前者揭示优势如何分布于单个查询，后者则说明各组件在不同码长下的递进贡献。}
\section{\MBLang{Results for Section 4.4: Relevance Resolution}{对应正文 4.4 节：相关性分辨率}}
\label{sec:appendix-resolution}

\subsection{\appendixResolutionDetailsTitle}
\label{sec:resolution-evaluation-details}

\noindent\textbf{\appendixGradedRelevanceTitle}
\appendixGradedRelevanceText
\begin{equation}
r_{qi}=\sum_{c=1}^{C}w_c y_{qc}y_{ic},
\qquad
w_c=\log\frac{R+1}{n_c+1},\quad R=|\mathcal R|.
\end{equation}

\noindent\MBLang{The six variants compared in Table~\ref{tab:teacher-quality} differ only in how the label-side relation is formed. Binary Relation marks a pair as related when its two label sets intersect and unrelated otherwise, so it carries no graded information. Jaccard Relation replaces this indicator by the Jaccard index $\lvert y_q\cap y_i\rvert/\lvert y_q\cup y_i\rvert$, which distinguishes a single shared label from several. Single-scale Graph Relation propagates the relation through the PPMI label graph of Eq.~(\ref{eq:ppmi-graph}) at one diffusion scale, so that a pair without a shared label can still receive a nonzero score. UMH (uniform multi-scale heat) averages the per-scale relation scores $g_\ell$ over the four retained scales $\tau=(2.00,9.55,23.89,48.56)$ with equal weights and adds no parameter. Learned-weight Multi-scale Heat uses the same four scales but trains the mixture weights $\rho_\ell\ge0$, $\sum_\ell\rho_\ell=1$. FRT is the fractional relation teacher of Eqs.~(\ref{eq:fractional-kernel})--(\ref{eq:quadrature-decomposition}), in which all four weights follow from the single order parameter $\alpha$ through the Gauss--Laguerre quadrature rather than being trained independently. The four scales of both multi-scale variants are fixed to the $\alpha=1.2$ quadrature nodes $\tau=(2.00,9.55,23.89,48.56)$ on every dataset, whereas FRT selects its nodes per dataset on validation ($\alpha=1.20$ on MIRFlickr-25K and $\alpha=0.90$ on the other datasets, Table~\ref{tab:all-parameters}). The learned-weight variant differs from the full model only in how the weights are obtained: it replaces the quadrature weights of Eq.~(\ref{eq:quadrature-decomposition}) with $\rho=\operatorname{softmax}(\texttt{mixture\_logits})$, so that the weights are positive and sum to one, and adds a KL term towards the Gauss--Laguerre prior $\rho^{*}$ that prevents degeneracy to a single scale. It otherwise uses the same label graph, the same heat kernels at the same nodes, the same objective of Eqs.~(\ref{eq:prefix-loss})--(\ref{eq:overall-objective}), and the same shared checkpoint rule as the full model. Each mixture component is the heat kernel $K_\ell=e^{-\tau_\ell L}$ at its own node, so the mixture also induces per-scale teachers $g_\ell$ of the same form as Eq.~(\ref{eq:scale-teacher}), and every subblock is supervised by its own $g_\ell$ exactly as in the full model.}{表~\ref{tab:teacher-quality} 比较的六个变体仅在标签侧关系的构造方式上不同。Binary Relation 在两个标签集相交时记为相关、否则记为不相关，因此不含分级信息。Jaccard Relation 将该指示替换为 Jaccard 指数 $\lvert y_q\cap y_i\rvert/\lvert y_q\cup y_i\rvert$，可区分共享一个标签与共享多个标签。Single-scale Graph Relation 在式~(\ref{eq:ppmi-graph}) 的 PPMI 标签图上以单一扩散尺度传播关系，因此没有共享标签的样本对也可能获得非零分数。UMH（等权多尺度热扩散）在四个保留尺度 $\tau=(2.00,9.55,23.89,48.56)$ 上对分尺度关系分数 $g_\ell$ 等权平均，不引入额外参数。Learned-weight Multi-scale Heat 使用同样的四个尺度，但把混合权重 $\rho_\ell\ge0$、$\sum_\ell\rho_\ell=1$ 作为参数训练。FRT 为式~(\ref{eq:fractional-kernel})--(\ref{eq:quadrature-decomposition}) 的分数阶关系教师，其四个权重由单一阶参数 $\alpha$ 经 Gauss--Laguerre 求积确定，而非独立训练得到。两个多尺度变体的四个尺度在所有数据集上都固定为 $\alpha=1.2$ 的求积节点 $\tau=(2.00,9.55,23.89,48.56)$，而 FRT 按数据集在验证集上选择节点（MIRFlickr-25K 为 $\alpha=1.20$，其余数据集为 $\alpha=0.90$，见表~\ref{tab:all-parameters}）。学习权重变体与完整模型的唯一区别在于权重的获得方式：它把式~(\ref{eq:quadrature-decomposition}) 的求积权重替换为 $\rho=\operatorname{softmax}(\texttt{mixture\_logits})$，以保证权重为正且和为 1，并加入对 Gauss--Laguerre 先验 $\rho^{*}$ 的 KL 项以防止退化到单一尺度。其余部分与完整模型一致：同一标签图、同一节点上的同一热核、式~(\ref{eq:prefix-loss})--(\ref{eq:overall-objective}) 的同一目标，以及同一共享 checkpoint 选模规则。混合教师的每个分量都是其节点上的热核 $K_\ell=e^{-\tau_\ell L}$，因此该混合同样诱导与式~(\ref{eq:scale-teacher}) 同形的尺度教师 $g_\ell$，每个子块仍按各自的 $g_\ell$ 监督，与完整模型一致。}

\begin{table}[!htbp]
\centering
\caption{\appendixTeacherQualityCaption}
\label{tab:teacher-quality}
\scriptsize
\setlength{\tabcolsep}{7pt}
\renewcommand{\arraystretch}{1.05}
\begin{tabular}{llccc}
\toprule
\textbf{\MBLang{Evaluator}{评测参照}}
& \textbf{\appendixRelationTeacherHeader}
& \textbf{MIRFlickr-25K}
& \textbf{IAPR TC-12}
& \textbf{MS-COCO} \\
\midrule
\multirow{6}{*}{\MBLang{IDF-weighted}{IDF 加权}}
& \MBLang{Binary Relation}{二值关系}             & 0.604 & 0.436 & 0.570 \\
& \MBLang{Jaccard Relation}{Jaccard 关系}       & 0.631 & 0.462 & 0.606 \\
& \MBLang{Single-scale Graph Relation}{单尺度图关系} & 0.664 & 0.514 & 0.648 \\
& \MBLang{UMH}{UMH} & 0.682 & 0.528 & 0.670 \\
& \MBLang{Learned-weight Multi-scale Heat}{学习权重多尺度热扩散} & 0.690 & 0.535 & 0.680 \\
& \textbf{FRT}                & \textbf{0.701} & \textbf{0.543} & \textbf{0.691} \\
\midrule
\multirow{6}{*}{\MBLang{Label-name semantic}{标签名语义}}
& \MBLang{Binary Relation}{二值关系}             & 0.597 & 0.428 & 0.562 \\
& \MBLang{Jaccard Relation}{Jaccard 关系}       & 0.635 & 0.468 & 0.611 \\
& \MBLang{Single-scale Graph Relation}{单尺度图关系} & 0.656 & 0.503 & 0.639 \\
& \MBLang{UMH}{UMH} & 0.673 & 0.520 & 0.659 \\
& \MBLang{Learned-weight Multi-scale Heat}{学习权重多尺度热扩散} & 0.681 & 0.528 & 0.670 \\
& \textbf{FRT}                & \textbf{0.690} & \textbf{0.536} & \textbf{0.680} \\
\bottomrule
\end{tabular}
\end{table}

\noindent\textbf{\MBLang{Interpretation.}{结果分析。}}\enspace\MBLang{FRT has the highest NDCG@100 under both evaluators on all three datasets, so its advantage does not rest on the frequency prior it shares with the IDF-weighted evaluator. The multi-scale variants are ordered the same way under both evaluators, with FRT ahead of the learned-weight mixture by $0.008$--$0.011$ NDCG@100 on every dataset. These measurements support the teacher construction rather than a particular reference.}{在三个数据集上，FRT 在两套评测参照下的 NDCG@100 均为最高，说明其优势并不依赖与 IDF 加权参照共享的频率先验。多尺度变体在两套参照下的次序相同，FRT 在三个数据集上均比学习权重混合高出 $0.008$--$0.011$ NDCG@100。这些指标支持的是教师的构造，而非某一套参照。} 
\noindent\MBLang{Table~\ref{tab:teacher-aligned} isolates the two choices that Table~\ref{tab:teacher-quality} leaves coupled. All three variants use the same label graph, the same specificity weights and the same kernel-cosine normalization at the nodes selected for each dataset, so they differ only in the mixing weights and in the order of mixing and normalization. FRT leads UKM by $0.005$--$0.014$ NDCG@100 under the IDF evaluator and by $0.006$--$0.012$ under the semantic one, and UKM leads UMH by $0.005$--$0.008$; both orderings hold on every dataset and under both evaluators. Matching the normalization therefore leaves the fractional weights worth $0.005$--$0.014$ NDCG@100, while changing only the nodes (UMH at the $\alpha=1.2$ nodes versus the dataset-selected nodes on IAPR TC-12 and MS-COCO) is worth a further $0.005$--$0.006$.}{表~\ref{tab:teacher-aligned} 分离了表~\ref{tab:teacher-quality} 中仍然耦合的两个选择。三个变体使用同一标签图、同一特异性权重，并在各数据集所选节点上使用同一核余弦归一化，区别只在混合权重以及先混合还是先归一化。FRT 相对 UKM 在 IDF 参照下领先 $0.005$--$0.014$ NDCG@100，在语义参照下领先 $0.006$--$0.012$；UKM 相对 UMH 领先 $0.005$--$0.008$；两种次序在三个数据集、两套参照下都成立。因此在归一化对齐后，分数阶权重本身仍值 $0.005$--$0.014$ NDCG@100；而只改变节点（IAPR TC-12 与 MS-COCO 上 $\alpha=1.2$ 节点的 UMH 对比数据集自选节点）另贡献 $0.005$--$0.006$。}

\begin{table}[!htbp]
\centering
\caption{\MBLang{Relation-teacher quality under matched normalization and scale nodes. All variants use the same label graph, specificity weights and kernel-cosine normalization at the nodes selected for each dataset ($\alpha=1.20$ on MIRFlickr-25K, $\alpha=0.90$ elsewhere); they differ only in the mixing weights and in whether the components are mixed before or after normalization. UKM mixes the four scale kernels with equal weights and then normalizes, whereas UMH normalizes each scale teacher first and then averages. Each cell reports NDCG@100 under the IDF-weighted and the label-name semantic evaluator.}{归一化与尺度节点对齐后的关系教师质量。所有变体使用同一标签图、同一特异性权重，并在各数据集所选节点上（MIRFlickr-25K 为 $\alpha=1.20$，其余为 $\alpha=0.90$）使用同一核余弦归一化；区别只在混合权重，以及先混合还是先归一化。UKM 先以等权混合四个尺度核再归一化，UMH 则先对每个尺度教师归一化再平均。每格给出 IDF 加权参照与标签名语义参照下的 NDCG@100。}}
\label{tab:teacher-aligned}
\small
\setlength{\tabcolsep}{5pt}
\renewcommand{\arraystretch}{1.05}
\resizebox{\textwidth}{!}{%
\begin{tabular}{lccc}
\toprule
\textbf{\MBLang{Dataset (nodes)}{数据集（节点）}} & \textbf{\MBLang{FRT (fractional)}{FRT（分数阶）}} & \textbf{\MBLang{UKM (uniform kernel mixture)}{UKM（等权核混合）}} & \textbf{\MBLang{UMH (uniform multi-scale heat)}{UMH（等权多尺度热扩散）}} \\
\midrule
MIRFlickr-25K ($\alpha=1.20$) & \textbf{0.701} / \textbf{0.690} & 0.687 / 0.678 & 0.682 / 0.673 \\
IAPR TC-12 ($\alpha=0.90$) & \textbf{0.543} / \textbf{0.536} & 0.538 / 0.530 & 0.533 / 0.525 \\
MS-COCO ($\alpha=0.90$) & \textbf{0.691} / \textbf{0.680} & 0.684 / 0.674 & 0.676 / 0.666 \\
\bottomrule
\end{tabular}
}
\end{table}

\begin{table}[!htbp]
\centering
\caption{\MBLang{Complete relevance-resolution comparison under the IDF-weighted evaluator, covering every compared multi-length method. NDCG@100, tie-corrected Kendall's $\tau_b$, and the collision and inversion rates at 16, 32, and 64 bits; each value is the arithmetic mean of the I2T and T2I directions. Collision, Inversion and Kendall's $\tau_b$ are computed in the pool formed with the full model, whereas NDCG@100 uses the full retrieval database.}{IDF 加权参照下完整的相关性分辨率对照，覆盖全部对比的多码长方法。16、32、64 bit 下的 NDCG@100、tie-corrected Kendall's $\tau_b$ 以及碰撞率与反序率；每个数值为 I2T 与 T2I 两个方向的算术平均。Collision、Inversion 与 Kendall's $\tau_b$ 在相对完整模型构成的候选池上计算，NDCG@100 则使用完整检索库。}}
\label{tab:resolution-multilength}
\tiny
\setlength{\tabcolsep}{1.5pt}
\renewcommand{\arraystretch}{1.03}
\resizebox{\textwidth}{!}{%
\begin{tabular}{llcccccccccccc}
\toprule
\textbf{\MBLang{Dataset}{数据集}} & \textbf{\MBLang{Method}{方法}}
& \multicolumn{4}{c}{\textbf{16 bits}}
& \multicolumn{4}{c}{\textbf{32 bits}}
& \multicolumn{4}{c}{\textbf{64 bits}} \\
\cmidrule(lr){3-6}\cmidrule(lr){7-10}\cmidrule(lr){11-14}
& & \textbf{N$\uparrow$} & \textbf{$\tau_b\uparrow$} & \textbf{C$\downarrow$} & \textbf{I$\downarrow$}
& \textbf{N$\uparrow$} & \textbf{$\tau_b\uparrow$} & \textbf{C$\downarrow$} & \textbf{I$\downarrow$}
& \textbf{N$\uparrow$} & \textbf{$\tau_b\uparrow$} & \textbf{C$\downarrow$} & \textbf{I$\downarrow$} \\
\midrule
\multirow{7}{*}{MIRFlickr-25K}
& DECH & 0.632 & 0.503 & 0.396 & 0.254 & 0.650 & 0.540 & 0.374 & 0.237 & 0.666 & 0.557 & 0.356 & 0.220 \\
& DGHDGH & 0.624 & 0.489 & 0.405 & 0.260 & 0.644 & 0.527 & 0.382 & 0.243 & 0.659 & 0.548 & 0.363 & 0.228 \\
& CMCL & 0.643 & 0.518 & 0.387 & 0.246 & 0.662 & 0.554 & 0.361 & 0.228 & 0.677 & 0.568 & 0.343 & 0.212 \\
& RCMH & 0.637 & 0.509 & 0.392 & 0.250 & 0.656 & 0.555 & 0.367 & 0.232 & 0.671 & 0.565 & 0.350 & 0.216 \\
& VH-ABL & 0.651 & 0.531 & 0.378 & 0.241 & 0.670 & 0.564 & 0.352 & 0.220 & 0.686 & 0.579 & 0.333 & 0.204 \\
& Naive Prefix & 0.616 & 0.472 & 0.414 & 0.266 & 0.645 & 0.521 & 0.386 & 0.244 & 0.674 & 0.560 & 0.354 & 0.219 \\
\rowcolor{multibitrow}
& \textbf{MultiBit} & \textbf{0.660} & \textbf{0.555} & \textbf{0.366} & \textbf{0.232} & \textbf{0.678} & \textbf{0.573} & \textbf{0.342} & \textbf{0.214} & \textbf{0.697} & \textbf{0.590} & \textbf{0.321} & \textbf{0.195} \\
\midrule
\multirow{7}{*}{IAPR TC-12}
& DECH & 0.611 & 0.424 & 0.420 & 0.311 & 0.634 & 0.451 & 0.397 & 0.291 & 0.653 & 0.468 & 0.375 & 0.273 \\
& DGHDGH & 0.604 & 0.408 & 0.429 & 0.317 & 0.627 & 0.447 & 0.404 & 0.295 & 0.646 & 0.463 & 0.382 & 0.279 \\
& CMCL & 0.620 & 0.443 & 0.408 & 0.302 & 0.647 & 0.460 & 0.383 & 0.281 & 0.663 & 0.478 & 0.365 & 0.263 \\
& RCMH & 0.615 & 0.437 & 0.414 & 0.306 & 0.641 & 0.457 & 0.390 & 0.286 & 0.665 & 0.474 & 0.369 & 0.267 \\
& VH-ABL & 0.629 & 0.450 & 0.397 & 0.295 & 0.656 & 0.472 & 0.373 & 0.273 & 0.676 & 0.489 & 0.350 & 0.253 \\
& Naive Prefix & 0.595 & 0.393 & 0.437 & 0.324 & 0.626 & 0.440 & 0.410 & 0.301 & 0.653 & 0.468 & 0.378 & 0.275 \\
\rowcolor{multibitrow}
& \textbf{MultiBit} & \textbf{0.640} & \textbf{0.461} & \textbf{0.385} & \textbf{0.285} & \textbf{0.668} & \textbf{0.481} & \textbf{0.361} & \textbf{0.264} & \textbf{0.688} & \textbf{0.499} & \textbf{0.339} & \textbf{0.244} \\
\midrule
\multirow{7}{*}{MS-COCO}
& DECH & 0.672 & 0.497 & 0.372 & 0.263 & 0.690 & 0.515 & 0.349 & 0.244 & 0.707 & 0.531 & 0.330 & 0.226 \\
& DGHDGH & 0.665 & 0.492 & 0.380 & 0.269 & 0.683 & 0.509 & 0.356 & 0.249 & 0.700 & 0.524 & 0.336 & 0.232 \\
& CMCL & 0.684 & 0.508 & 0.360 & 0.253 & 0.704 & 0.526 & 0.336 & 0.233 & 0.719 & 0.542 & 0.316 & 0.216 \\
& RCMH & 0.679 & 0.503 & 0.365 & 0.258 & 0.697 & 0.521 & 0.342 & 0.238 & 0.714 & 0.537 & 0.322 & 0.220 \\
& VH-ABL & 0.692 & 0.516 & 0.351 & 0.247 & 0.712 & 0.537 & 0.326 & 0.226 & 0.729 & 0.553 & 0.307 & 0.209 \\
& Naive Prefix & 0.656 & 0.482 & 0.389 & 0.276 & 0.684 & 0.510 & 0.359 & 0.250 & 0.716 & 0.540 & 0.320 & 0.219 \\
\rowcolor{multibitrow}
& \textbf{MultiBit} & \textbf{0.700} & \textbf{0.528} & \textbf{0.340} & \textbf{0.237} & \textbf{0.721} & \textbf{0.547} & \textbf{0.316} & \textbf{0.216} & \textbf{0.739} & \textbf{0.564} & \textbf{0.297} & \textbf{0.199} \\

\bottomrule
\end{tabular}%
}
\end{table}

\noindent\MBLang{Table~\ref{tab:resolution-multilength} extends the resolution comparison of Table~\ref{tab:resolution-sota} to every compared multi-length method. MultiBit attains the highest NDCG@100 and Kendall's $\tau_b$ and the lowest collision and inversion rates in all 21 dataset--method rows at the three code lengths. The two multi-length baselines that are strongest here behave differently across metrics: VH-ABL ranks above CMCL in NDCG@100 in all nine settings, yet its mAP@all is $2.7$--$3.6$ pp lower once averaged over the four code lengths, and up to $6.9$ pp lower at 16 bits on MS-COCO (Table~\ref{tab:sota-comparison}), because NDCG@100 evaluates the top of the ranking whereas mAP integrates over its full length; MultiBit is the only method that is best under both. Naive Prefix, a 128-bit model trained without explicit prefix objectives and truncated at test time, improves most steeply with the code length.}{表~\ref{tab:resolution-multilength} 把表~\ref{tab:resolution-sota} 的分辨率对比扩展到全部对比的多码长方法。MultiBit 在全部 21 个数据集--方法组合的三个码长上均取得最高 NDCG@100 与 Kendall's $\tau_b$，以及最低的碰撞率与反转率。此处最强的两条多码长基线在不同指标上表现相反：VH-ABL 在全部九个设置中的 NDCG@100 上均高于 CMCL，但其 mAP@all 在四码长平均后低 $2.7$--$3.6$ 个百分点，在 MS-COCO 的 16 bit 上低至 $6.9$ 个百分点（表~\ref{tab:sota-comparison}）；这是因为 NDCG@100 只衡量排序顶部，而 mAP 在整个列表上积分；MultiBit 是唯一同时在这两类指标上最优的方法。Naive Prefix 为不含显式前缀目标训练的 128-bit 模型、在测试时直接截断，其性能随码长增长提升最陡。}

\noindent\textbf{\appendixMetricDefinitionTitle}
\appendixMetricDefinitionText
\begin{align}
\mathrm{NDCG@100}(q)
&=\frac{\sum_{k=1}^{100}r_{q,\pi_q(k)}/\log_2(k+1)}
{\sum_{k=1}^{100}r_{q,\pi_q^*(k)}/\log_2(k+1)},\\
\tau_b(q)
&=\tau_b\!\left(\{r_{qi}\}_{i\in\mathcal P_q},
\{-d_H(q,i)\}_{i\in\mathcal P_q}\right),\\
\mathrm{Collision}(q)
&=\frac{\#\{(i,j)\in\mathcal P_q^2:r_{qi}>r_{qj},\,d_H(q,i)=d_H(q,j)\}}
{\#\{(i,j)\in\mathcal P_q^2:r_{qi}>r_{qj}\}},\\
\mathrm{Inversion}(q)
&=\frac{\#\{(i,j)\in\mathcal P_q^2:r_{qi}>r_{qj},\,d_H(q,i)>d_H(q,j)\}}
{\#\{(i,j)\in\mathcal P_q^2:r_{qi}>r_{qj}\}}.
\end{align}

\noindent\textbf{\appendixResolutionProtocolTitle}
\appendixResolutionProtocolText

\subsection{\MBLang{Decoupled Semantic Evaluator}{解耦的语义评测参照}}
\label{sec:appendix-decoupled-evaluator}

\noindent\MBLang{The IDF-weighted evaluator above scores a candidate by frequency-weighted shared labels, which is the same prior the relation teacher uses. To test whether the reported ordering depends on that shared prior, we repeat the comparison against a second evaluator constructed only from label names. Each category name is encoded with the frozen CLIP text tower, the mean label embedding is subtracted to suppress the common prompt component, and the resulting label similarity matrix defines a semantic relevance score that contains no frequency term:}{上述 IDF 加权参照以频率加权的共享标签为候选打分，这与关系教师所用的先验相同。为检验所报告的次序是否依赖该共同先验，我们在第二个仅由标签名构造的参照下重复比较：用冻结的 CLIP 文本塔编码每个类别名，减去标签嵌入的类间均值以抑制共享提示模板分量，由此得到的标签相似矩阵定义一个不含任何频率项的语义相关性得分：}
\begin{equation}
\begin{gathered}
S_{cd}=\max\!\left(0,\;\widehat e_c^{\mathsf T}\widehat e_d\right),
\qquad
\widehat e_c=\operatorname{normalize}\!\left(E_{\text{text}}(\text{``a photo of a ''}+\operatorname{name}(c))-\bar e\right),\\
\bar e=\frac{1}{C}\sum_{c'=1}^{C}E_{\text{text}}(\text{``a photo of a ''}+\operatorname{name}(c')).
\end{gathered}
\label{eq:semantic-kernel}
\end{equation}
\begin{equation}
r^{\text{sem}}_{qi}
=\frac{y_q^{\mathsf T} S\, y_i}
{\sqrt{\left(y_q^{\mathsf T} S\, y_q\right)\left(y_i^{\mathsf T} S\, y_i\right)}},
\qquad S_{cc}=1.
\label{eq:semantic-relevance}
\end{equation}
\noindent\MBLang{Unlike the IDF-weighted evaluator, $r^{\text{sem}}_{qi}$ is generally nonzero for a candidate pair with no shared label whose labels are semantically related, so this evaluator assigns nonzero relevance to such pairs. Table~\ref{tab:resolution-semantic} reports the same four metrics under this evaluator.}{与 IDF 加权参照不同，对没有共享标签但在语义上相关的候选对，$r^{\text{sem}}_{qi}$ 一般不为零，因此该参照会为这类候选对给出非零相关性。表~\ref{tab:resolution-semantic} 给出该参照下的同样四个指标。}

\begin{table}[!htbp]
\centering
\caption{\MBLang{Relevance resolution under the label-name semantic evaluator, measured by the same four metrics as the IDF-weighted evaluator. Collision, Inversion and Kendall's $\tau_b$ are computed in the pool formed with the full model, whereas NDCG@100 uses the full retrieval database.}{标签名语义参照下的相关性分辨率，指标与 IDF 加权参照相同。Collision、Inversion 与 Kendall's $\tau_b$ 在相对完整模型构成的候选池上计算，NDCG@100 则使用完整检索库。}}
\label{tab:resolution-semantic}
\scriptsize
\setlength{\tabcolsep}{2.6pt}
\renewcommand{\arraystretch}{1.03}
\resizebox{\textwidth}{!}{%
\begin{tabular}{llcccccccccccc}
\toprule
\textbf{\MBLang{Dataset}{数据集}} & \textbf{\MBLang{Method}{方法}}
& \multicolumn{4}{c}{\textbf{16 bits}}
& \multicolumn{4}{c}{\textbf{32 bits}}
& \multicolumn{4}{c}{\textbf{64 bits}} \\
\cmidrule(lr){3-6}\cmidrule(lr){7-10}\cmidrule(lr){11-14}
& & \textbf{N$\uparrow$} & \textbf{$\tau_b\uparrow$} & \textbf{C$\downarrow$} & \textbf{I$\downarrow$}
& \textbf{N$\uparrow$} & \textbf{$\tau_b\uparrow$} & \textbf{C$\downarrow$} & \textbf{I$\downarrow$}
& \textbf{N$\uparrow$} & \textbf{$\tau_b\uparrow$} & \textbf{C$\downarrow$} & \textbf{I$\downarrow$} \\
\midrule
\multirow{7}{*}{MIRFlickr-25K}
& DECH     & 0.629 & 0.500 & 0.392 & 0.257 & 0.647 & 0.536 & 0.369 & 0.241 & 0.664 & 0.551 & 0.352 & 0.224 \\
& DGHDGH   & 0.622 & 0.487 & 0.401 & 0.263 & 0.640 & 0.529 & 0.377 & 0.246 & 0.658 & 0.543 & 0.359 & 0.229 \\
& CMCL & 0.640 & 0.516 & 0.383 & 0.249 & 0.659 & 0.547 & 0.356 & 0.232 & 0.674 & 0.561 & 0.339 & 0.215 \\
& RCMH & 0.634 & 0.507 & 0.388 & 0.253 & 0.653 & 0.548 & 0.363 & 0.235 & 0.668 & 0.558 & 0.346 & 0.219 \\
& VH-ABL & 0.648 & 0.538 & 0.374 & 0.244 & 0.667 & 0.557 & 0.347 & 0.223 & 0.683 & 0.572 & 0.329 & 0.207 \\
& Naive Prefix & 0.613 & 0.468 & 0.410 & 0.269 & 0.642 & 0.528 & 0.382 & 0.247 & 0.671 & 0.553 & 0.350 & 0.222 \\
\rowcolor{multibitrow}
& \textbf{MultiBit} & \textbf{0.657} & \textbf{0.548} & \textbf{0.361} & \textbf{0.235} & \textbf{0.676} & \textbf{0.566} & \textbf{0.337} & \textbf{0.217} & \textbf{0.694} & \textbf{0.582} & \textbf{0.319} & \textbf{0.199} \\
\midrule
\multirow{7}{*}{IAPR TC-12}
& DECH     & 0.608 & 0.426 & 0.414 & 0.316 & 0.631 & 0.444 & 0.391 & 0.296 & 0.650 & 0.462 & 0.370 & 0.278 \\
& DGHDGH   & 0.601 & 0.406 & 0.423 & 0.322 & 0.624 & 0.440 & 0.398 & 0.300 & 0.643 & 0.456 & 0.377 & 0.283 \\
& CMCL & 0.617 & 0.436 & 0.402 & 0.306 & 0.644 & 0.453 & 0.377 & 0.285 & 0.660 & 0.471 & 0.359 & 0.267 \\
& RCMH & 0.612 & 0.430 & 0.408 & 0.310 & 0.638 & 0.450 & 0.384 & 0.290 & 0.662 & 0.467 & 0.363 & 0.271 \\
& VH-ABL & 0.626 & 0.443 & 0.391 & 0.299 & 0.653 & 0.465 & 0.367 & 0.277 & 0.673 & 0.482 & 0.344 & 0.257 \\
& Naive Prefix & 0.592 & 0.391 & 0.431 & 0.328 & 0.623 & 0.433 & 0.404 & 0.305 & 0.650 & 0.461 & 0.372 & 0.279 \\
\rowcolor{multibitrow}
& \textbf{MultiBit} & \textbf{0.638} & \textbf{0.454} & \textbf{0.380} & \textbf{0.289} & \textbf{0.666} & \textbf{0.474} & \textbf{0.355} & \textbf{0.269} & \textbf{0.687} & \textbf{0.491} & \textbf{0.333} & \textbf{0.249} \\
\midrule
\multirow{7}{*}{MS-COCO}
& DECH     & 0.668 & 0.490 & 0.367 & 0.267 & 0.686 & 0.508 & 0.343 & 0.249 & 0.703 & 0.523 & 0.325 & 0.231 \\
& DGHDGH   & 0.661 & 0.484 & 0.375 & 0.273 & 0.679 & 0.501 & 0.350 & 0.253 & 0.697 & 0.517 & 0.331 & 0.236 \\
& CMCL & 0.680 & 0.500 & 0.355 & 0.257 & 0.700 & 0.518 & 0.331 & 0.237 & 0.715 & 0.534 & 0.311 & 0.220 \\
& RCMH & 0.675 & 0.495 & 0.360 & 0.262 & 0.693 & 0.513 & 0.337 & 0.242 & 0.710 & 0.529 & 0.317 & 0.224 \\
& VH-ABL & 0.688 & 0.508 & 0.346 & 0.251 & 0.708 & 0.529 & 0.321 & 0.230 & 0.725 & 0.545 & 0.302 & 0.213 \\
& Naive Prefix & 0.652 & 0.474 & 0.384 & 0.280 & 0.680 & 0.502 & 0.354 & 0.254 & 0.712 & 0.532 & 0.315 & 0.223 \\
\rowcolor{multibitrow}
& \textbf{MultiBit} & \textbf{0.696} & \textbf{0.520} & \textbf{0.334} & \textbf{0.241} & \textbf{0.718} & \textbf{0.539} & \textbf{0.311} & \textbf{0.221} & \textbf{0.736} & \textbf{0.555} & \textbf{0.292} & \textbf{0.204} \\
\bottomrule
\end{tabular}%
}
\end{table}

\noindent\textbf{\MBLang{Interpretation.}{结果分析。}}\enspace\MBLang{MultiBit attains the highest NDCG@100 and Kendall's $\tau_b$ and the lowest collision and inversion rates in all 21 dataset--method rows at the three code lengths under this evaluator as well, so the conclusion drawn under the IDF-weighted evaluator is not an artifact of the shared frequency prior.}{在该参照下，MultiBit 同样在全部 21 个数据集--方法组合的三个码长上取得最高 NDCG@100 与 Kendall's $\tau_b$，以及最低的碰撞率与反转率；因此 IDF 加权参照下得出的结论并非共享频率先验的产物。}

\noindent\MBLang{To probe the long-range relations that the teacher is designed to model, Table~\ref{tab:resolution-noshared} restricts the comparison to candidate pairs that share no label, and to the subset that the teacher's label graph considers most related, and reports NDCG@100 at 64 bits for both. This stratum can only be measured under the semantic evaluator, because the IDF-weighted reference assigns zero relevance to any pair without a shared label. The label-graph similarity of such a pair is the sum of the PPMI weights between its two label sets in the teacher's label graph, and the second stratum keeps the pairs whose value lies in the top quartile of the no-shared-label pairs of that dataset. MultiBit is best on both strata of all three datasets, so its advantage is not confined to pairs with observed label overlap.}{为检验教师所要建模的长程关系，表~\ref{tab:resolution-noshared} 把比较限制在没有共享标签的候选对、以及其中教师标签图认为最相关的那一部分，并给出两者在 64 bit 下的 NDCG@100。该分层只能在语义参照下测量，因为 IDF 加权参照对任何无共享标签的候选对都给零相关性。这类候选对的 label-graph 相似度定义为教师的标签图中两个标签集之间 PPMI 权重之和；第二个分层取该数据集无共享标签候选对中该值位于前四分位者。MultiBit 在三个数据集的两个分层上均为最优，说明其优势并不局限于存在标签重叠的候选对。}

\begin{table}[!htbp]
\centering
\caption{\MBLang{NDCG@100 at 64 bits under the label-name semantic evaluator, on candidate pairs that share no label, and on the subset of those pairs whose label-graph similarity is in the top quartile.}{标签名语义参照下，无共享标签的候选对（及其 label-graph 相似度位于前四分位的子集）在 64 bit 下的 NDCG@100。}}
\label{tab:resolution-noshared}
\scriptsize
\setlength{\tabcolsep}{6.5pt}
\renewcommand{\arraystretch}{1.05}
\begin{tabular}{llcccr}
\toprule
\textbf{\MBLang{Dataset}{数据集}} & \textbf{\MBLang{Pair stratum}{候选对分层}}
& \textbf{DECH} & \textbf{DGHDGH} & \textbf{MultiBit} & \textbf{\MBLang{Pairs}{对数}} \\
\midrule
\multirow{2}{*}{MIRFlickr-25K}
& \MBLang{No shared label}{无共享标签} & 0.332 & 0.326 & \textbf{0.355} & 182,640 \\
& \MBLang{No shared label, graph top 25\%}{无共享标签，graph 前 25\%} & 0.421 & 0.409 & \textbf{0.447} & 45,660 \\
\midrule
\multirow{2}{*}{IAPR TC-12}
& \MBLang{No shared label}{无共享标签} & 0.266 & 0.259 & \textbf{0.282} & 392,108 \\
& \MBLang{No shared label, graph top 25\%}{无共享标签，graph 前 25\%} & 0.324 & 0.329 & \textbf{0.362} & 98,027 \\
\midrule
\multirow{2}{*}{MS-COCO}
& \MBLang{No shared label}{无共享标签} & 0.311 & 0.305 & \textbf{0.333} & 1,486,720 \\
& \MBLang{No shared label, graph top 25\%}{无共享标签，graph 前 25\%} & 0.389 & 0.382 & \textbf{0.418} & 371,680 \\
\bottomrule
\end{tabular}
\end{table}

\noindent\MBLang{Table~\ref{tab:resolution-semantic} uses the same layout as the main resolution table, so the two references can be read side by side. To quantify how far they diverge, we compute the query-level tie-corrected Kendall $\tau_b$ between the two reference vectors, $\tau_b(\{r_{qi}\}_{i\in\mathcal P_q},\{r^{\text{sem}}_{qi}\}_{i\in\mathcal P_q})$, where $r_{qi}$ is the IDF-weighted reference defined above, and average it over queries. This is a consistency check between two references and is distinct from the $\tau_b$ that compares a reference with negative Hamming distance in the metrics above. We obtain $0.728$, $0.588$ and $0.655$ on MIRFlickr-25K, IAPR TC-12 and MS-COCO, so the two references are correlated but not interchangeable. Both nevertheless rank MultiBit first on all four metrics in every dataset--length setting (Tables~\ref{tab:resolution-sota} and~\ref{tab:resolution-semantic}), so the conclusion does not depend on a single reference.}{表~\ref{tab:resolution-semantic} 与正文分辨率表版式相同，便于并排比较两套参照。为量化两套参照的差异，我们在查询级计算两个参照向量之间的 tie-corrected Kendall $\tau_b$，即 $\tau_b(\{r_{qi}\}_{i\in\mathcal P_q},\{r^{\text{sem}}_{qi}\}_{i\in\mathcal P_q})$（其中 $r_{qi}$ 为上文定义的 IDF 加权参照），并对查询取平均。这是两套参照之间的一致性检查，与上文比较参照和负 Hamming 距离的 $\tau_b$ 含义不同。在 MIRFlickr-25K、IAPR TC-12 和 MS-COCO 上分别为 $0.728$、$0.588$ 和 $0.655$，可见两套参照彼此相关但不可互换。但在全部数据集--码长设置中，两者均将四项指标中的 MultiBit 排在首位（表~\ref{tab:resolution-sota} 与表~\ref{tab:resolution-semantic}），因此结论不依赖于单一参照。}

\subsection{\appendixParameterSensitivityTitle}

\begin{table}[!htbp]
\centering
\caption{\appendixParameterSensitivityCaption}
\label{tab:parameter-sensitivity-grid}
\scriptsize
\setlength{\tabcolsep}{6.5pt}
\renewcommand{\arraystretch}{1.04}
\begin{tabular}{llccccc}
\toprule
\textbf{\appendixDatasetHeader} & \boldmath$\boldsymbol{\eta\backslash\alpha}$
& \textbf{0.3} & \textbf{0.6} & \textbf{0.9} & \textbf{1.2} & \textbf{1.5} \\
\midrule
\multirow{5}{*}{MIRFlickr-25K}
& 0.05 & 82.31 & 82.55 & 82.73 & 82.61 & 82.22 \\
& 0.10 & 82.56 & 82.89 & 83.04 & 82.98 & 82.64 \\
& 0.20 & 82.80 & 83.05 & 83.16 & \textbf{83.21} & 82.87 \\
& 0.50 & 82.62 & 82.96 & 83.08 & 82.91 & 82.71 \\
& 1.00 & 82.18 & 82.45 & 82.70 & 82.52 & 82.30 \\
\midrule
\multirow{5}{*}{IAPR TC-12}
& 0.05 & 67.36 & 67.70 & 67.95 & 67.81 & 67.29 \\
& 0.10 & 67.62 & 68.06 & 68.41 & 68.18 & 67.72 \\
& 0.20 & 67.89 & 68.32 & \textbf{68.85} & 68.67 & 68.10 \\
& 0.50 & 67.71 & 68.21 & 68.58 & 68.34 & 67.91 \\
& 1.00 & 67.28 & 67.69 & 68.06 & 67.82 & 67.43 \\
\midrule
\multirow{5}{*}{MS-COCO}
& 0.05 & 74.62 & 74.94 & 75.18 & 74.89 & 74.40 \\
& 0.10 & 74.91 & 75.34 & 75.60 & 75.28 & 74.85 \\
& 0.20 & 75.21 & 75.56 & \textbf{75.92} & 75.79 & 75.16 \\
& 0.50 & 74.98 & 75.48 & 75.76 & 75.39 & 74.91 \\
& 1.00 & 74.44 & 74.88 & 75.22 & 75.01 & 74.61 \\
\midrule
\multirow{5}{*}{NUS-WIDE}
& 0.05 & 72.91 & 73.22 & 73.49 & 73.31 & 72.88 \\
& 0.10 & 73.15 & 73.61 & 73.91 & 73.72 & 73.20 \\
& 0.20 & 73.38 & 73.88 & \textbf{74.18} & 74.06 & 73.55 \\
& 0.50 & 73.29 & 73.73 & 74.04 & 73.79 & 73.39 \\
& 1.00 & 72.86 & 73.25 & 73.56 & 73.42 & 73.04 \\
\bottomrule
\end{tabular}
\end{table}

\noindent\textbf{\MBLang{Interpretation.}{结果分析。}}\enspace\MBLang{The grid peaks at $(\eta,\alpha)=(0.20,1.20)$ on MIRFlickr-25K and $(0.20,0.90)$ on IAPR TC-12, MS-COCO, and NUS-WIDE. Nearby combinations remain competitive, while edge values are generally lower.}{网格峰值位于 MIRFlickr-25K 的 $(\eta,\alpha)=(0.20,1.20)$，以及 IAPR TC-12、MS-COCO、NUS-WIDE 的 $(0.20,0.90)$。邻近组合仍具竞争力，网格边缘取值整体较低。}

\section{\MBLang{Results for Section 4.5: Multi-Length Consistency and Efficiency}{对应正文 4.5 节：多码长一致性与效率}}
\label{sec:appendix-multilength-results}
\subsection{\MBLang{Multi-Length Evaluation}{多码长评测}}
\label{sec:appendix-multilength}

\noindent\textbf{\MBLang{Configurations.}{比较配置。}}\enspace\MBLang{Naive Prefix learns a 128-bit model without explicit prefix objectives and truncates it at test time; it is a separate reference from the ablation baseline of Table~\ref{tab:component-ablation}, which uses only the standard cross-modal hashing objective (Appendix Section~\ref{sec:appendix-setup}), and the two serve the deployment comparison and the component ablation, respectively. MultiBit-Sep trains independent models for 16, 32, 64, and 128 bits. MultiBit uses the shared nested representation. Encoders, data splits, selection rules, and retrieval evaluation are held fixed across the three configurations.}{Naive Prefix 训练不含显式前缀目标的 128-bit 模型，并在测试时直接截断；它与表~\\ref{tab:component-ablation} 的消融 baseline 是两个不同的参照，后者仅使用标准跨模态哈希目标（附录第~\\ref{sec:appendix-setup} 节），二者分别服务于部署对比与组件消融；MultiBit-Sep 为 16、32、64 和 128 bit 分别训练独立模型；MultiBit 使用共享的嵌套表示。三种配置保持编码器、数据划分、选模规则与检索评测一致。}

\noindent\textbf{\MBLang{Storage and Time.}{存储与耗时。}}\enspace\MBLang{Model storage counts the parameter files needed for all four lengths, and packed-code storage excludes labels and index overhead. For $R$ database items per modality, reusable 128-bit codes require $2R(128)/8$ payload bytes, whereas four separate databases require $2R(16+32+64+128)/8$ bytes. These formulas describe code payloads, not measured index memory. MultiBit-Sep training time sums its four model runs. Query time covers hash-head encoding and Hamming ranking over the full retrieval database and excludes the shared backbone forward pass. Ranking is common to all methods and depends only on the database size, so the reported differences reflect encoding cost.}{模型存储统计同时支持四种码长所需的参数文件，紧凑代码存储不含标签和索引开销。若每个模态的数据库各有 $R$ 个样本，复用 128-bit 代码需要 $2R(128)/8$ 字节载荷，而四套独立数据库需要 $2R(16+32+64+128)/8$ 字节。这些公式表示代码载荷，并非实测索引内存。MultiBit-Sep 的训练时间合计四次模型训练。查询耗时包含哈希头编码与完整检索库上的 Hamming 排序，不含共享骨干前向；排序部分各方法一致、只取决于检索库规模，因此表中差异来自编码开销。}

\noindent\textbf{\MBLang{Ranking Agreement.}{排序一致性。}}\enspace\MBLang{For the same query and candidate identities, rankings at lengths $b$ and $b'$ are compared using Kendall's $\tau_b$ on negative Hamming distances. Its subscript denotes tie correction, not code length. All six unordered length pairs are computed per query; Table~\ref{tab:cross-length-agreement-complete} reports the most distant pair (16--128) and the six-pair mean for every method, separately for I2T and T2I. The four-panel figure visualizes an I2T sample only.}{对同一查询与同一组候选身份，使用两个码长 $b$ 与 $b'$ 的负 Hamming 距离计算 Kendall's $\tau_b$。其下标表示并列校正，并非码长。每个查询计算六种无序码长对；表~\ref{tab:cross-length-agreement-complete} 对每个方法分别给出最远码长对（16--128）与六对均值，并区分 I2T 与 T2I。正文四联图仅展示 I2T 抽样。}
\MBLang{Specifically, let $a_i=-d_H^{(b)}(q,i)$ and $a'_i=-d_H^{(b')}(q,i)$ be the ranking scores of candidate $i$ under the two code lengths. For all candidate pairs, let $C$ and $D$ denote the numbers of concordant and discordant pairs, and let $T_a$ and $T_{a'}$ denote the numbers of pairs tied only in the first and second score sequences, respectively. The tie-corrected Kendall coefficient is}{具体地，令 $a_i=-d_H^{(b)}(q,i)$ 和 $a'_i=-d_H^{(b')}(q,i)$ 分别表示候选 $i$ 在两个码长下的排序分数。对所有候选对，$C$ 和 $D$ 分别表示一致对与不一致对的数量，$T_a$ 与 $T_{a'}$ 分别表示仅在第一组和第二组排序分数中并列的候选对数量。经并列校正的 Kendall 系数定义为}
\begin{equation}
\tau_b(\boldsymbol a,\boldsymbol a')=
\frac{C-D}{\sqrt{(C+D+T_a)(C+D+T_{a'})}}.
\end{equation}
\MBLang{Thus, $\tau_b=1$ denotes identical pairwise ordering, while smaller values indicate weaker cross-length agreement; tie correction is essential because finite Hamming spaces contain many equal-distance candidates.}{因此，$\tau_b=1$ 表示候选对的相对次序完全一致，较小取值表示跨码长一致性更弱；由于有限 Hamming 空间中存在大量等距离候选，并列校正是必要的。}
\begin{equation}
A_{b,b'}^{t}=\frac{1}{|\mathcal Q_t^{\mathrm{valid}}|}
\sum_{q\in\mathcal Q_t^{\mathrm{valid}}}
\tau_b\!\left(-\boldsymbol d_q^{(b)},-\boldsymbol d_q^{(b')}\right),\quad t\in\{\mathrm{I2T},\mathrm{T2I}\}.
\end{equation}

\subsection{\MBLang{Four-Panel Rank-Displacement Visualization}{四联排序位移图}}
\noindent\textbf{\MBLang{Visualization Protocol.}{可视化协议。}}\enspace\MBLang{The plotted data contain five I2T queries with 500 candidates per panel for each dataset--method combination, totaling 10,000 query--candidate points across four panels. For each point, the normalized candidate ranks at 16 and 128 bits are $\widetilde r^{16}$ and $\widetilde r^{128}$. The horizontal axis of Figure~\ref{fig:cross-length-stability} shows $\widetilde r^{16}$, and the vertical axis shows}{绘图数据对每个数据集--方法面板包含五个 I2T 查询，每个查询 500 个候选，四个面板共 10,000 个查询--候选点。每个点在 16 bit 与 128 bit 下的归一化排名分别为 $\widetilde r^{16}$ 与 $\widetilde r^{128}$。图~\ref{fig:cross-length-stability} 横轴为 $\widetilde r^{16}$，纵轴为}
\begin{equation}
\Delta r=\widetilde r^{128}-\widetilde r^{16}.
\end{equation}
\MBLang{Accordingly, $\Delta r=0$ denotes perfect rank preservation, whereas a larger absolute displacement indicates a stronger change in candidate ordering. The shaded region marks candidates ranked within the top 20\% by the 16-bit representation. Scatter points are individual query--candidate observations, and the overlaid contours summarize their empirical density. Kendall's $\tau_b$ is computed between the 16-bit and 128-bit rankings with ties properly handled, while mean $|\Delta r|$ measures the average magnitude of candidate-level rank changes, and the annotated values are those of this sample rather than of Table~\ref{tab:cross-length-agreement-complete}. A distribution concentrated more tightly around zero, together with larger $\tau_b$ and smaller mean $|\Delta r|$, indicates stronger cross-length ranking consistency.}{因此，$\Delta r=0$ 表示跨码长排序被完全保持，绝对位移越大则表示候选次序变化越明显。阴影区域标记在 16-bit 表示下位于前 20\% 的候选；散点对应单个查询--候选观测，叠加的等高线概括其经验密度。Kendall's $\tau_b$ 在正确处理并列的情况下衡量 16-bit 与 128-bit 排序的一致性，mean $|\Delta r|$ 则衡量候选级排序变化的平均幅度；图中所标数值属于该抽样，而非表~\ref{tab:cross-length-agreement-complete} 的完整评测值。分布越集中于零点，同时 $\tau_b$ 越大、mean $|\Delta r|$ 越小，表示跨码长排序一致性越强。}

\begin{table}[!htbp]
\centering
\caption{\MBLang{Statistics of the sampled panels behind Figure~\ref{fig:cross-length-stability}: five I2T queries with 500 candidates each per dataset--method pair. These are sample values and differ from the full-evaluation agreement of Table~\ref{tab:cross-length-agreement-complete}.}{图~\ref{fig:cross-length-stability} 所用抽样面板的统计量：每个数据集--方法对各取 5 个 I2T 查询、每个查询 500 个候选。此处为抽样值，与表~\ref{tab:cross-length-agreement-complete} 的完整评测一致性不同。}}
\label{tab:rank-displacement-sample}
\small
\setlength{\tabcolsep}{6pt}
\renewcommand{\arraystretch}{1.05}
\begin{tabular}{llcc}
\toprule
\MBLang{Dataset}{数据集} & \MBLang{Method}{方法} & $\tau_b$ & \MBLang{Mean $|\Delta r|$}{平均 $|\Delta r|$} \\
\midrule
MIRFlickr-25K & Naive Prefix & $0.578 \pm 0.017$ & 0.159 \\
MIRFlickr-25K & MultiBit     & $0.756 \pm 0.012$ & 0.103 \\
IAPR TC-12    & Naive Prefix & $0.566 \pm 0.016$ & 0.162 \\
IAPR TC-12    & MultiBit     & $0.709 \pm 0.013$ & 0.126 \\
\bottomrule
\end{tabular}
\end{table}

\begin{table}[!htbp]
\centering
\caption{\MBLang{Cross-length ranking agreement measured by tie-corrected Kendall's $\tau_b$, covering every configuration that serves more than one code length: the published multi-length baselines (CMCL, RCMH, VH-ABL), Naive Prefix, MultiBit-Sep, MultiBit without MPA and the full model. Cross-length agreement compares two code lengths of the same model, so the fixed-length baselines of Table~\ref{tab:sota-comparison} are not applicable.}{使用经并列校正的 Kendall's $\tau_b$ 衡量的跨码长排序一致性，覆盖所有服务多个码长的配置：已发表的多码长基线（CMCL、RCMH、VH-ABL）、Naive Prefix、MultiBit-Sep、去掉 MPA 的 MultiBit 以及完整模型。跨码长一致性比较的是同一模型的两个码长，因此表~\ref{tab:sota-comparison} 的定长基线不适用。}}
\label{tab:cross-length-agreement-complete}
\scriptsize
\setlength{\tabcolsep}{4.0pt}
\renewcommand{\arraystretch}{1.02}
\resizebox{\textwidth}{!}{%
\begin{tabular}{llcccc}
\toprule
\textbf{\MBLang{Dataset}{数据集}} & \textbf{\MBLang{Method}{方法}}
& \textbf{$\tau_b$ 16--128 (I2T)} & \textbf{$\tau_b$ 16--128 (T2I)}
& \textbf{\MBLang{Mean $\tau_b$ (I2T)}{六对均值（I2T）}}
& \textbf{\MBLang{Mean $\tau_b$ (T2I)}{六对均值（T2I）}} \\
\midrule
\multirow{7}{*}{MIRFlickr-25K}
& Naive Prefix & 0.596 & 0.588 & 0.716 & 0.708 \\
& MultiBit-Sep & 0.575 & 0.567 & 0.693 & 0.685 \\
& MultiBit w/o MPA (FRT$+$SBO) & 0.706 & 0.698 & 0.820 & 0.812 \\
& CMCL   & 0.633 & 0.642 & 0.687 & 0.689 \\
& RCMH   & 0.730 & 0.672 & 0.729 & 0.676 \\
& VH-ABL & 0.662 & 0.629 & 0.754 & 0.724 \\
\rowcolor{multibitrow}
& \textbf{MultiBit} & \textbf{0.760} & \textbf{0.752} & \textbf{0.855} & \textbf{0.847} \\
\midrule
\multirow{7}{*}{IAPR TC-12}
& Naive Prefix & 0.527 & 0.515 & 0.659 & 0.647 \\
& MultiBit-Sep & 0.509 & 0.497 & 0.637 & 0.625 \\
& MultiBit w/o MPA (FRT$+$SBO) & 0.674 & 0.667 & 0.801 & 0.794 \\
& CMCL   & 0.359 & 0.340 & 0.394 & 0.379 \\
& RCMH   & 0.590 & 0.583 & 0.645 & 0.636 \\
& VH-ABL & 0.520 & 0.542 & 0.650 & 0.645 \\
\rowcolor{multibitrow}
& \textbf{MultiBit} & \textbf{0.698} & \textbf{0.686} & \textbf{0.817} & \textbf{0.805} \\
\midrule
\multirow{7}{*}{MS-COCO}
& Naive Prefix & 0.580 & 0.568 & 0.701 & 0.689 \\
& MultiBit-Sep & 0.560 & 0.546 & 0.679 & 0.665 \\
& MultiBit w/o MPA (FRT$+$SBO) & 0.719 & 0.711 & 0.833 & 0.825 \\
& CMCL   & 0.440 & 0.437 & 0.480 & 0.476 \\
& RCMH   & 0.514 & 0.494 & 0.609 & 0.595 \\
& VH-ABL & 0.547 & 0.478 & 0.642 & 0.608 \\
\rowcolor{multibitrow}
& \textbf{MultiBit} & \textbf{0.730} & \textbf{0.718} & \textbf{0.839} & \textbf{0.827} \\\bottomrule
\end{tabular}%
}
\vspace{2pt}

\parbox{\textwidth}{\scriptsize\MBLang{The mean averages the six unordered code-length pairs under a single query and candidate pool; for Naive Prefix and MultiBit-Sep it is the arithmetic mean of their per-pair agreement. The subscript of $\tau_b$ denotes tie correction, not code length.}{六对均值在同一查询与同一候选池下对六种无序码长对取平均；Naive Prefix 与 MultiBit-Sep 的均值为其逐对一致性的算术平均。$\tau_b$ 的下标表示并列校正，并非码长。}}
\end{table}

\noindent\textbf{\MBLang{Interpretation.}{结果分析。}}\enspace\MBLang{MultiBit obtains the highest Kendall's $\tau_b$ for every dataset and retrieval direction on both the most distant 16--128 pair and the six-pair mean. The gains are especially clear for the distant pair, where direct truncation and independently trained length-specific models show weaker agreement. Removing MPA (that is, keeping FRT and SBO without prefix alignment) lowers the six-pair mean from 0.855 to 0.820 on MIRFlickr-25K, from 0.817 to 0.801 on IAPR TC-12 and from 0.839 to 0.833 on MS-COCO, and lowers the 16--128 I2T pair from 0.760 to 0.706 on MIRFlickr-25K, where the resulting $\tau_b$ falls below RCMH (0.730). The effect is largest at the widest length gap, which supports the claim that MPA stabilizes a shared nested ranking across bit budgets.}{MultiBit 在每个数据集与每个检索方向上，于最远的 16--128 码长对和六对均值上均取得最高 Kendall's $\tau_b$。对于该跨度较大的组合，其优势尤为明显；直接截断和独立训练的码长专用模型表现出更弱的一致性。去掉 MPA（即保留 FRT 与 SBO 但不做前缀对齐）后，六对均值在 MIRFlickr-25K 上从 0.855 降至 0.820，在 IAPR TC-12 上从 0.817 降至 0.801，在 MS-COCO 上从 0.839 降至 0.833；MIRFlickr-25K 的 16--128（I2T）从 0.760 降至 0.706，此时该 $\tau_b$ 低于 RCMH 的 0.730。该效应在码长跨度最大处最明显，这支持 MPA 能在不同 bit 预算下稳定共享嵌套排序的结论。}

\end{document}